\documentclass[aps,prb,amsmath,amssymb,amsfont,showpacs,superscriptaddress,reprint]{revtex4-1}
\usepackage{amsmath}
\usepackage{bm}
\usepackage[dvipsnames]{xcolor}
\usepackage{braket}
\usepackage[artemisia]{textgreek}  
\usepackage{float}        
\usepackage{graphicx}
\usepackage{times}

\DeclareMathOperator{\expe}{e}
\newcommand{\cplxi}{\mathrm{i}}
\newcommand{\dd}{\mathrm{d}}

\begin{document}

\title{Electronic properties of air-sensitive nanomaterials probed with microwave impedance measurements}

\author{B. G. M\'{a}rkus}
\affiliation{Department of Physics, Budapest University of Technology and Economics and MTA-BME Lend\"{u}let Spintronics Research Group (PROSPIN), P.O. Box 91, H-1521 Budapest, Hungary}

\author{G. Cs\H{o}sz}
\affiliation{Department of Physics, Budapest University of Technology and Economics and MTA-BME Lend\"{u}let Spintronics Research Group (PROSPIN), P.O. Box 91, H-1521 Budapest, Hungary}

\author{O. S\'{a}gi}
\affiliation{Department of Physics, Budapest University of Technology and Economics and MTA-BME Lend\"{u}let Spintronics Research Group (PROSPIN), P.O. Box 91, H-1521 Budapest, Hungary}

\author{B. Gy\"{u}re-Garami}
\affiliation{Department of Physics, Budapest University of Technology and Economics and MTA-BME Lend\"{u}let Spintronics Research Group (PROSPIN), P.O. Box 91, H-1521 Budapest, Hungary}

\author{V. Lloret}
\affiliation{Department of Chemistry and Pharmacy Friedrich-Alexander-Universit\"{a}t Erlangen-N\"{u}rnberg (FAU) D-91054 Erlangen, Germany and Joint Institute of Advanced Materials and Processes (ZMP), D-90762 F\"{u}rth, Germany}

\author{S. Wild}
\affiliation{Department of Chemistry and Pharmacy Friedrich-Alexander-Universit\"{a}t Erlangen-N\"{u}rnberg (FAU) D-91054 Erlangen, Germany and Joint Institute of Advanced Materials and Processes (ZMP), D-90762 F\"{u}rth, Germany}

\author{G. Abell\'{a}n}
\affiliation{Department of Chemistry and Pharmacy Friedrich-Alexander-Universit\"{a}t Erlangen-N\"{u}rnberg (FAU) D-91054 Erlangen, Germany and Joint Institute of Advanced Materials and Processes (ZMP), D-90762 F\"{u}rth, Germany}

\author{N. M. Nemes}
\affiliation{GFMC, Unidad Asociada ICMM-CSIC "Laboratorio de Heteroestructuras con Aplicacion en Espintronica", Departamento de Fisica de Materiales Universidad Complutense de Madrid, 28040 Madrid, Spain}

\author{G. Klupp}
\affiliation{Institute for Solid State Physics and Optics, Wigner Research Centre for Physics, Hungarian Academy of Sciences, P.O. Box 49, H-1525 Budapest, Hungary}

\author{K. Kamar\'{a}s}
\affiliation{Institute for Solid State Physics and Optics, Wigner Research Centre for Physics, Hungarian Academy of Sciences, P.O. Box 49, H-1525 Budapest, Hungary}

\author{A. Hirsch}
\affiliation{Department of Chemistry and Pharmacy Friedrich-Alexander-Universit\"{a}t Erlangen-N\"{u}rnberg (FAU) D-91054 Erlangen, Germany and Joint Institute of Advanced Materials and Processes (ZMP), D-90762 F\"{u}rth, Germany}

\author{F. Hauke}
\affiliation{Department of Chemistry and Pharmacy Friedrich-Alexander-Universit\"{a}t Erlangen-N\"{u}rnberg (FAU) D-91054 Erlangen, Germany and Joint Institute of Advanced Materials and Processes (ZMP), D-90762 F\"{u}rth, Germany}

\author{F. Simon}
\thanks{Corresponding author: \textsf{f.simon@eik.bme.hu}, Phone: +36-1-463-1215, Fax: +36-1-463-4180}
\affiliation{Department of Physics, Budapest University of Technology and Economics and MTA-BME Lend\"{u}let Spintronics Research Group (PROSPIN), P.O. Box 91, H-1521 Budapest, Hungary}


\begin{abstract}
  Characterization of electronic properties of novel materials is of great importance for exploratory materials development and also for the discovery of new correlated phases. As several novel compounds are available in powder form only, contactless methods, which also work on air-sensitive samples, are highly desired. We present that the microwave cavity perturbation technique is a versatile tool to study conductivity in such systems. The examples include studies on semiconducting-metallic crossover in carbon nanotubes upon alkali doping, study of vortex motion in the K$_3$C$_{60}$ superconductor, and the characterization of various alkali atom doped phases of black phosphorus.
\end{abstract}

\maketitle

\section{Introduction}
Research of novel materials is in the forefront of contemporary physics and chemistry due to the presence of fundamentally interesting phases and the broad range of applications. In particular low-dimensional nanomaterials, including fullerenes \cite{Kroto1985}, carbon nanotubes (CNT) \cite{Iijima1991}, graphene \cite{Novoselov2004}. and most recently black phosphorus \cite{Li2014,Liu2014,castellanos-gomez2015,ling2015,Liu2015,Ryder2016,gusmao2017}, attracted significant attention. Common to these materials is that the electronic properties can be finely tuned by charge transfer which led to e.g. the discovery of superconductivity \cite{Hebard1991}, spin-density waves \cite{Bommeli1995}, and a Mott transition \cite{Maeda2002} in fullerides, the bleaching of optical transitions \cite{Eklund1997} and the Tomonaga--Luttinger to Fermi liquid crossover \cite{Pichler2004} in single walled-carbon nanotubes (SWCNT). It is also common that alkali atom doped modifications of these materials are extremely air-sensitive thus conventional contact probing of the electronic properties is difficult. This difficulty is even more pronounced for black phosphorus (bP) for which the pristine, undoped material is highly air and moisture sensitive \cite{ziletti2015,AbellanJACS2017,ZhangJACS2018}.

In principle, contactless methods, such as infrared spectroscopy could provide the required information on the conductivity of such samples, {\color{black}especially when combined with ESR, NMR or Raman-spectroscopy}. The microwave frequency range arises as an automatic choice for these studies as the frequency is closer to DC. As a result, this type of measurements is more representative for the DC conductivity except for some exotic cases e.g. including heavy fermions \cite{Dressel2005,Dressel2013}. Microwave cavity perturbation \cite{Buravov1971,Klein1993} allows to determine \emph{relative} changes in the conductivity, $\sigma$, dielectric permittivity, $\epsilon_{\text{r}}$, and magnetic permeability, $\mu_{\text{r}}$, of air-sensitive nanomaterials: the samples are placed inside a microwave cavity and changes in the the quality factor, $Q$, and resonance frequency, $f$, allow to determine the relevant parameters. A disadvantage of the method is that absolute values of material parameters are difficult to attain. An important advantage of the method is that the appropriate choice of cavity resonance mode allows to disentangle the different properties, e.g. there are cavity modes which locally sustain a node in the electric or magnetic field.

To illustrate the versatility of this technique we present experiments on pristine and potassium intercalated black phosphorus (bP) showing the appearance of new metallic charge carriers. Afterwards the K$_3$C$_{60}$ samples are investigated near the superconducting phase transition, which is followed by \emph{in-situ} doping of single walled carbon nanotubes (SWCNTs), whose resistivity evolution is examined in real time. {\color{black}Furthermore the technique was used for many other materials previously, such as: superconductivity in Nb and Pb \cite{HolczerPRB}; insulating behavior of NH$_3$K$_3$C$_{60}$ systems \cite{MaedaPRL}; and many more \cite{DresselReview}.}

\section{Experimental}

In this section first we present briefly how the investigated samples are prepared and introduce the reader to microwave conductivity measurements.

\paragraph{Sample preparation}

Polycrystalline black phosphorus was purchased from smart-elements with purity of $99.998\%$. The crystals were ground inside an argon-filled glovebox (MBraun GmbH with $< 0.1$ ppm of O$_2$ and H$_2$O) and used as a crushed powder. Potassium intercalation of the material was carried out in a solid-state route: to the pulverized phosphorus the stoichiometric amount of potassium metal (Sigma-Aldrich, $99.95\%$) was added to obtain the required intercalation compound (in our case $1:8$ for KP$_8$). Afterwards, the mixture was heated up to $70~^{\circ}$C, where it was carefully mixed together. The detailed process of synthesis is described in a previous article \cite{Abellan2017}. For the experiments, $10.1$ and $10.3$ mg of bP and KP$_8$ was put into ESR grade quartz tubes, evacuated to high vacuum ($2 \times 10^{-6}$ mbar) and sealed under $20$ mbar of He ($6.0$ Helium), which allowed cryogenic measurements.

Single crystal and powder K$_3$C$_{60}$ samples were prepared by the conventional potassium intercalation method; the crystal sample was from the same batch as in Ref. \cite{NemesPRB2000}. The powder samples were further ground together with non-conducting SnO$_2$ powder to prevent conducting links between the grains. SQUID magnetometry attested that DC superconducting properties (such as the steepness of the superconducting transition) were unaffected by the mixing. Samples were sealed in quartz ampules under low pressure helium.

We used commercial SWCNTs prepared by the arc-discharge method. The material was obtained from Nanocarblab (Moscow, Russia) with a well known mean diameter of $d=1.4$ nm and variance of $\sigma_d=0.1$ nm. The diameter distribution can be estimated well by a Gaussian function. This batch is the same as we used previously during Raman \cite{SimonPRB2005}, NMR \cite{SimonPRL2005} and ESR \cite{SzirmaiPRB2017} characterization and peapod filling \cite{SimonPRL2006}. The material was purified with oxidation in air and various acid treatments \cite{SimonChemPhysLett2004}. Afterwards it was ground thoroughly to allow microwaves to penetrate the whole bulk, and to assist the intercalation of potassium. About $5$ mg of SWCNT material was then vacuum annealed at $500~^{\circ}$C for $1$ hour in an ESR quartz sample tube with a neck and inserted into an Ar filled glove-box without air exposure. To achieve potassium intercalation we followed the well developed path of two chamber vapor phase intercalation \cite{Dresselhaus1981} adopted to our setup. The K was placed above the nanotube powder, where the neck forbid the mixing of the materials, similarly to our previous work done on graphite \cite{FabianPRB2012}. Finally the sample was evacuated to vacuum without air exposure.

\paragraph{Microwave cavity perturbation technique}

Microwave properties were measured with the cavity perturbation method \cite{Klein1993,Donovan1993} as a function of temperature, $T$, and for the case of K$_3$C$_{60}$ samples various static magnetic fields, $B$, were also applied. The used copper cavity has an unloaded quality factor of $Q_0 \approx 10,000$ and a resonance frequency, $f_0 \approx 11.2$ GHz, whose temperature dependence is taken into account. The samples were placed in the node of the microwave electric field and maximum of the microwave magnetic field inside the TE011 cavity, which is the appropriate geometry to study minute changes in the conductivity \cite{KitanoPRL2002}. The alternating microwave magnetic field induces eddy currents in the sample, which causes a change in the microwave loss and shifts the resonator frequency. The $Q$ factor of the cavity is measured via rapid frequency sweeps near the resonance. A fit to the obtained resonance curve yields the position, $f$, and width, $\Gamma$, of the resonance. $Q$ is afterwards obtained from its definition $Q=f/\Gamma$. This value has to be corrected with the unloaded $Q$ factor of the cavity, thus the loss caused by the inserted sample is: $\mathrm{\Delta}\left(\frac{1}{2Q} \right) = \frac{1}{2Q}-\frac{1}{2Q_0}$ and the frequency shift is $\mathrm{\Delta}f/f_0 = (f-f_0)/f_0$. Physical properties such as the relative permittivity can be calculated in the following way:
\begin{equation}
\epsilon_{\text{r}}'-1 = - A \frac{V_{\text{s}}}{V_{\text{c}}}\times \mathrm{\Delta}f/f_0,
\end{equation}
and
\begin{equation}
\epsilon_{\text{r}}''=A \frac{V_{\text{s}}}{V_{\text{c}}} \times \mathrm{\Delta}\left(\frac{1}{2Q} \right),
\end{equation}
with $\epsilon_{\text{r}}=\epsilon_{\text{r}}' + \cplxi \epsilon_{\text{r}}''$. $V_{\text{s}}$ and $V_{\text{c}}$ are denoting the volume of the sample and the cavity, respectively. The $A$ constant can be calculated from the electric field in the cavity with, $E$, and without, $E_0$, the sample:
\begin{equation}
A\approx\frac{\int \dd V E_{0}^{\ast}E}{\int \dd V \left| E_0 \right|^2}.
\end{equation}
Calculation of the resistivity or conductivity from the measured parameters is summarized in the following section and in Refs \cite{Csosz2018,Gyure2018}.

For the black phosphorus and the fulleride samples, we applied a low temperature geometry as the interesting physics is expected to happen below room temperature. To achieve this, we placed our probehead into the cryostat of a superconducting magnet fabricated by Cryogenics Ltd. The temperature range of the setup can be varied between $3.3$ K up to $200$ K with magnetic field ranging from $0$ to $9$ T. The static magnetic field and the rf magnetic field are parallel in our geometry. The trapped flux of the solenoid is about $10-20$ mT or $100-200$ Oe. For the K$_3$C$_{60}$ samples zero field measurements were done in another cryostat without a magnet, where the temperature range is limited to $6.5$ K from below. The used low temperature setup is presented in details in Ref. \cite{KarsaPssb2012}.

In case of \emph{in-situ} measurements, high temperature is mandatory, for this a different setup is used with a very similar microwave cavity. The samples, which are sealed in a quartz capillary, are placed in an additional quartz insert, inside the cavity. Dry nitrogen gas flows through the insert, whose temperature can be varied between $100$ K and $1000$ K. The sample and the intercalant are placed in the same horizontal plane in our geometry to avoid molten potassium flowing inside the cavity. Outside of the insert the cavity is attached to a nitrogen purge line to remove the humidity and water, which would disturb the experiments due to their additional absorption. The temperature of the copper cavity is stabilized by water cooling to avoid thermal expansion. This setup could also serve to detect the opposite effect, e.g. defunctionalization or dedoping of materials, especially combined with e.g. Raman or TG-MS spectroscopy.

\section{Results}

We demonstrate the utility of microwave cavity measurements on the nowadays intensively studied low dimensional materials, such as black phosphorus ($2$D), single walled carbon nanotubes ($1$D) and fullerides ($0$D). In this work the conductivity of the materials was studied, however the method itself is not limited to this value, e.g. dielectric properties ($\epsilon_{\text{r}}$) can be examined. The first subsection is about black phosphorus and its alkali intercalated variant, namely KP$_8$. The second one concerns the superconducting phase transition in K$_3$C$_{60}$, emphasizing that the method is a great tool to study such materials, especially when it is combined with magnetic field. These samples were measured in the low temperature setup, in turn the method can also be combined with high temperature in a different geometry, enabling real time examination of vapor-phase intercalation. This is demonstrated in the third part on SWCNTs.

In microwave conductivity measurements, the sample morphology greatly affects the relation between the complex conductivity of the material, $\widetilde{\sigma}$, and the microwave parameters, the loss and shift. Two limiting cases are known: i) the sample is large and the field penetrates only into a limited distance from the surface. This approximates the measurement performed on the K$_3$C$_{60}$ single crystal. ii) The sample consists of small grains whose dimensions are comparable to the penetration depth, $\delta$. This approximates well the black phosphorus, the SWCNT and the powder K$_3$C$_{60}$ samples of divided small grains approximated with spherical particles \cite{Klein1993,Donovan1993,Csosz2018}.

In the first case, when the radio frequency field penetrates in the skin depth only, also referred as to the skin limit, the following equation holds between the microwave measurement parameters and the properties of the material:
\begin{equation}
\frac{\mathrm{\Delta}f}{f_0} - \cplxi \mathrm{\Delta} \left(\frac{1}{2Q} \right)=-\cplxi \nu Z_{\text{s}},
\end{equation}
where $Z_{\text{s}}$ is the surface impedance of the sample, related to the conductivity as $Z_\text{s} = \sqrt{\mu_\text{0}\omega/\cplxi\widetilde{\sigma}}$, $\widetilde{\sigma}=\sigma_1 + \cplxi \sigma_2$. {\color{black}Here $\sigma_1$ and $\sigma_2$ represents the real and the imaginary parts of the complex conductivity}. The dimensionless $\nu \ll 1$ parameter is the so-called resonator constant \cite{HolczerPRB1994} and it depends on the sample surface relative to that of the cavity. This is the most appropriate case for bulky samples, like single crystals.

In the second case, when the microwave field penetrates into the sample (known as the penetration limit), the cavity measurables depend differently on the sample parameters. It was shown for a sphere with a radius of $a$, that
\begin{gather}
\frac{\mathrm{\Delta}f}{f_0} - \cplxi \mathrm{\Delta} \left(\frac{1}{2Q} \right) = - \gamma \widetilde{\alpha}, \label{eq:pow} \\
\widetilde{\alpha} = \frac{1}{10} \left(a\widetilde{k}\right)^2,
\end{gather}
where $\widetilde{k}=\frac{\omega}{c} \sqrt{\cplxi \widetilde{\sigma}/\epsilon_0 \omega}$ is the complex wavenumber and $\gamma$ is a dimensionless sample volume dependent constant. This scenario is well applicable to finely ground powders. Further approximation can be made if the bulk conductivity and the particle sizes are known. In the case of bP and SWCNTs the $Q \sim \rho$ approximation holds.

\subsection{Tuning the electronic properties of black phosphorus}

\begin{figure}[h!]
\includegraphics*[width=.92\linewidth]{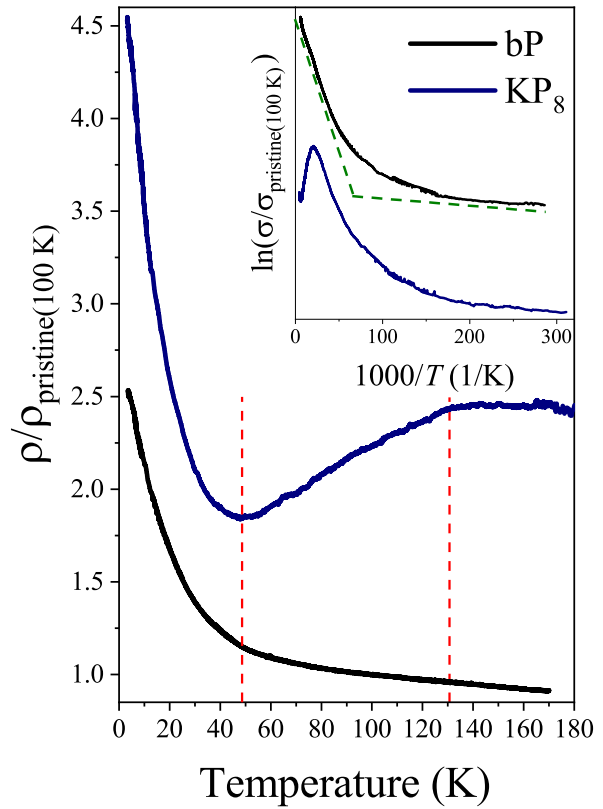}
\caption{Microwave conductivity measurement performed on pristine black phosphorus and its potassium intercalated derivative, KP$_8$. The pristine material exhibits a semiconducting behavior in the investigated temperature range, in contrast to the doped sample, that behaves in {\color{black}this} way below $49$ K and above $130$ K. The conductivity of this regime for both materials is dominated by the electrons excited thermally through the smaller gap. In KP$_8$ between $49$ and $130$ K a metallic regime is observed (noted with vertical dashed lines) yielding the presence of conduction electrons, similarly to our previous observations on the NaP$_6$ system \cite{Shiozawa2018}. We argue that the excess charges are present on the intercalated phosphorene sheets. Above $130$ K the resistivity is dominated again by the thermally excited electrons from the larger band gap. Please note that the resistivity values are normalized to the value of the pristine material taken at $100$ K. Furthermore the superconducting phase transition occurring at $3.8$ K \cite{Shiozawa2018,ZhangNatComm2017} is probably hindered by {\color{black}freezing} out of the thermal excitations as the volume fraction of the superconducting phase is low compared to the whole volume. {\color{black}Inset presents the observed data in an Arrhenius plot to emphasize the activated behavior and to make the metallic regime more pronounced. Dashed green lines are guides for the eye to demonstrate the two gaps present in the pristine material as observed in Ref. \cite{MarkusPssb2017}}.}
\label{fig:kp8}
\end{figure}

Conductivity measurements performed on undoped and potassium doped black phosphorus are shown in Fig. \ref{fig:kp8}. within the temperature range of $3.3$ K and $180$ K. The data points are normalized to the value of the pristine material taken at $100$ K (as our method cannot provide absolute values, this is a necessary step). The normalization was done taking into account the mass differences and assuming that the grain distribution of the material is not changed significantly during the intercalation. It is also assumed that $Q\sim \rho$ still holds for the KP$_8$ sample, as the conductivity of the sample is probably not increased by more than $3$ orders of magnitude.

The pristine material behaves as a semiconductor with a smaller and a larger band gap {\color{black}(visualized in the inset of Fig. \ref{fig:kp8})}, in agreement with previous literature observations \cite{MarkusPssb2017,Narita1983,Baba1991_1}. KP$_8$ exhibits a similar role below $49$ K with a slightly different small gap. The conductivity of this regime is dominated by the thermally excited electrons through the small band gap. The major difference between the two materials is visible in the $49-130$ K temperature range, where the alkali intercalated material exhibits a metallic behavior. {\color{black}It is clear that the intercalated system cannot be understood with only $2$ band gaps.} Similarly to our previous findings in the sodium-phosphorus system, NaP$_6$, we assign this to the presence of conduction electrons, whose states are existing in the intercalated phosphorene sheets \cite{Shiozawa2018}. Above $130$ K the conductivity of KP$_8$ is again dominated by the thermally excited electrons, but originated from the larger gap. Here we would like to point out that the overall resistivity of the intercalated material is increased by about a factor of $2$ compared to the pristine material. This somewhat strange and unexpected observation can be explained taking into account that during our measurements the microwave absorption is the quantity, which is truly measured and black phosphorus itself is a surprisingly good microwave absorbent \cite{MarkusPssb2017}. Moreover during the intercalation process some of the P$-$P bonds are broken \cite{Abellan2017}, especially near the surface, which can result in the suppression of absorption and increase of the resistivity. 

It was shown previously that alkali intercalated black phosphorus becomes superconductive at a universal transition temperature of ca. $3.8$ K \cite{Shiozawa2018,ZhangNatComm2017}. In our measurement this effect is hindered by {\color{black}freezing} out of the thermal excitations as the resistivity is increasing as the temperature is going to zero. This also means that the number of the {\color{black}conduction} electrons is low compared to the whole system, in agreement with the low superconducting volume fraction observed in SQUID measurements \cite{Shiozawa2018}.

\subsection{Superconductivity of K$_3$C$_{60}$}

\begin{figure}[h!]
\includegraphics*[width=\linewidth]{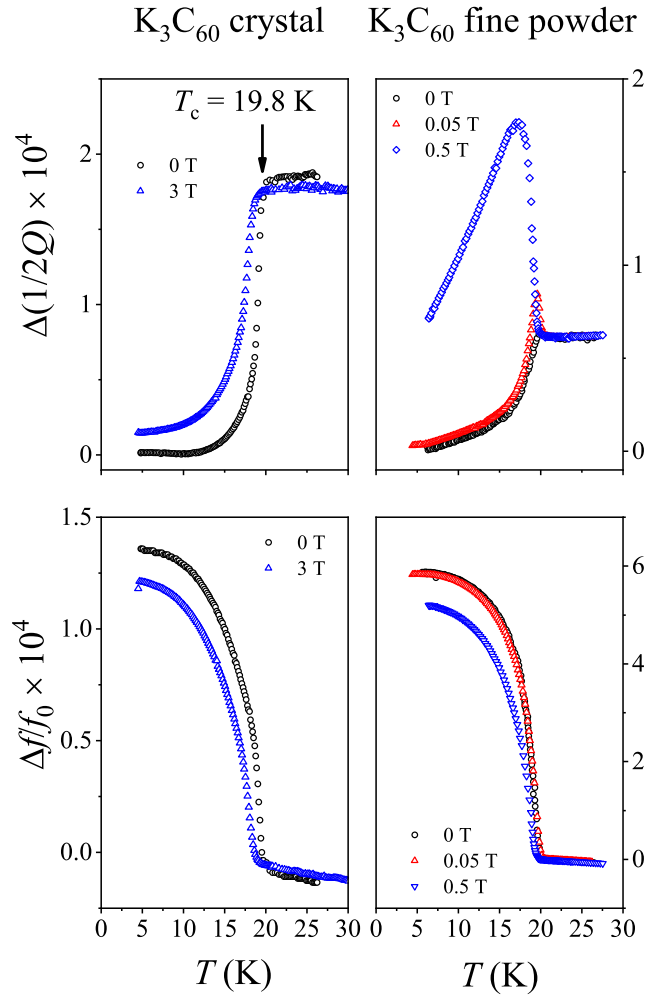}
\caption{Temperature dependent cavity loss, $\mathrm{\Delta}(1/2Q)$, and cavity frequency shift, $\mathrm{\Delta}f/f_0$ measurements for single crystal and powder K$_3$C$_{60}$ samples. The magnetic field was $0$ and $3$ T for the crystalline and $0$, $0.05$ and $0.5$ T for the powder sample. Note that the $Q$-factor changes significantly for the powder sample in contrast to the single crystal sample. Note the different scales for the $\mathrm{\Delta}f/f_0$ data. The obtained datasets can be fitted with theoretical curves calculated from the Coffey--Clem theory \cite{Csosz2018,CoffeyClemPRL1991,CoffeyClemPRB1992,CoffeyClemPRB19922,CoffeyClem19924,CoffeyClem1993}.}
\label{fig:k3c60}
\end{figure}

Fig. \ref{fig:k3c60} shows the microwave cavity loss, $\mathrm{\Delta}(1/2Q)$ and cavity shift, $\mathrm{\Delta}f/f_0$ for a single crystal and a fine powder K$_3$C$_{60}$ sample as a function of temperature for a few magnetic field values. The microwave loss decreases rapidly below $T_{\text{c}}$ in zero magnetic field as expected for superconductors. The most important observation is that the microwave loss becomes significant for a magnetic field as small as $0.05$ T for the fine powder sample, whereas even $3$ T has small effect on the microwave absorption for the single crystal one. In fact, we observe a huge, about $3$ times larger, microwave absorption below $T_{\text{c}}$ than in the normal state.

The fact that the enhanced microwave absorption occurs with the application of the magnetic field hints at a flux motion related phenomenon that is discussed in the framework of the Coffey--Clem (CC) theory \cite{CoffeyClemPRL1991,CoffeyClemPRB1992,CoffeyClemPRB19922,CoffeyClem19924,CoffeyClem1993}. The microwave absorption peak occurs above the irreversibility line, i.e. it is related to the physical behavior of the vortex-fluid state; for K$_3$C$_{60}$ $T_{\text{irrev}}(B = 0.1~\text{T}) \approx 15$ K and $T_{\text{irrev}}(B = 1~\text{T}) < 5$ K \cite{PrassidesBook}. Superconducting fullerides are type-II ($\lambda \gg \xi$) and have a short mean free path i.e. they can be described in the local electrodynamics limit, which simplifies the discussion \cite{GunnarsonRMP1997}. The phenomenological CC theory is based on a two-fluid model and considers the motion of vortices due to the exciting electromagnetic field in the presence of a viscous background (described by the viscous drag coefficient, $\eta$) and a restoring force (described by an effective pinning force constant, $\kappa_{\text{p}}$). The details of the theory and the exact calculation for the K$_3$C$_{60}$ described elsewhere \cite{Csosz2018}, here we only summarize the key points. The concept of the complex penetration depth, $\widetilde{\lambda}$ is introduced in the theory the following way:
\begin{equation}
\widetilde{\lambda}^2 = \frac{\lambda^2+(\cplxi/2)\widetilde{\delta}^2_{\text{eff}}}{1-2\cplxi \lambda^2/\widetilde{\delta}^2_{\text{nf}}},
\end{equation}
where $\widetilde{\delta}^2_{\text{nf}}$ is the skin depth in the normal fluid, $\lambda$ is the usual penetration depth and $\widetilde{\delta}^2_{\text{eff}}$ is the complex effective skin depth, which contains the effect of vortex motion. $\widetilde{\lambda}$ is linked to the conductivity by $\widetilde{\sigma} = \cplxi/\mu_0 \omega \widetilde{\lambda}^2$.

On the left panels of Fig. \ref{fig:k3c60} the application of the first case is shown, described above for the single crystal sample. We find that for both the calculation and experiment, the cavity loss parameter drops rapidly below $T_{\text{c}}$, although $\sigma_1/\sigma_{\text{n}}$, {\color{black}where $\sigma_{\text{n}}$ is the conductivity of the normal, Fermi liquid state,} is around unity due to the vortex motion. This effect is due to the development of a significant $\sigma_2/\sigma_{\text{n}} \sim 100$, which limits the penetration of microwaves into the sample and thus reduces the loss. This means that the microwave surface impedance measurement is less sensitive to provide information about $\sigma_1$ in the presence of vortex motion. 

The right panels in Fig. \ref{fig:k3c60} show the measurements for the fine powder sample. {\color{black}The arisen peak alike structure can be interpreted as follows: due to the vortex motion, the magnetic field can penetrate into a greater volume, than without, which results a decrease in the weight of Dirac delta function present in the real part of the conductivity. The effect is proportional to the inverse of the penetration depth squared. The missing weight, due to sum rule, results a constant part up to cut in frequency. If the vortex motion is not negligible, and the frequency used during the measurement is below this cut it can happen that the real part of the conductivity is higher in the superconducting state, than in the normal, Fermi-liquid state. This results an enhancement in the microwave loss, measured by the applied microwave technique.} To obtain fits, the transport and magnetic parameters ($\rho_{\text{n}}=1/\sigma_\text{n},\xi_0,\lambda_0$) have to be fixed to the respective mean of literature values. {\color{black}Here $\xi_0$ and $\lambda_0$ denotes the coherence length and the penetration depth at zero temperature, respectively.} We assumed that the sample consists of small spheres with a uniform diameter of $a$. The zero magnetic field data depend only on $\gamma$ and $a$ when the other sample properties are fixed. A fit to the $B = 0$ data yields $\gamma = 5.5(2) \times 10^{-4}$, and $a = 6.2(2)$ {\textmu}m. We then proceed to fit the magnetic field dependent data with $\kappa_{\text{p}}$ as the only free parameter and we obtain $\kappa_{\text{p}} = 1.0(1) \times 10^{-3}$ N$/$m$^2$. The fits and the detailed calculations for the obtained data are presented elsewhere \cite{Csosz2018}.

\subsection{\emph{In-situ} intercalation of potassium into SWCNT bundles; transition from semiconducting to metallic}

\begin{figure}[h!]
\includegraphics*[width=.85\linewidth]{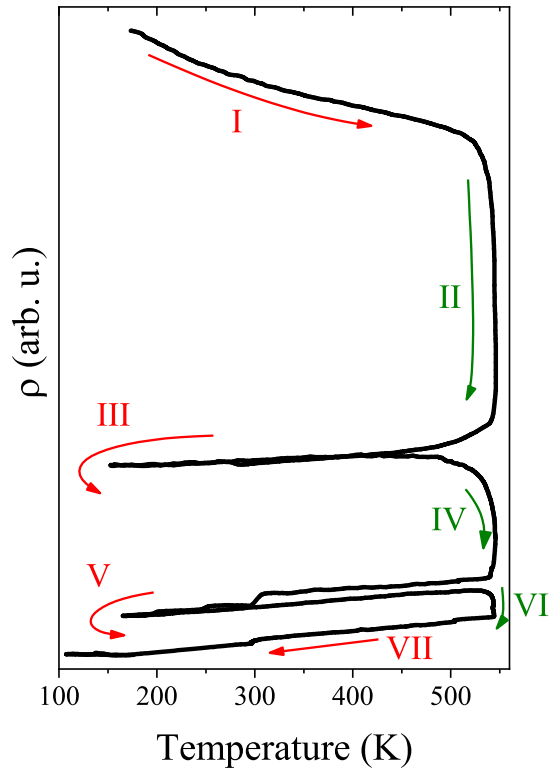}
\caption{\emph{In-situ} intercalation of arc discharge SWCNTs. Steps marked with {\color{black}red} color indicate measurement steps, while {\color{OliveGreen}green} parts show intercalation steps proceeded at $545$ K. In the {\color{black}first step} the material clearly shows a semiconducting behavior (resistivity decreases upon increasing temperature) as expected from a bundled branch of nanotubes, where $2/3$ of the tubes are non-metallic. In the {\color{OliveGreen}second step} the first intercalation takes place, which lasted for $16$ minutes. This is followed by the measurement denoted with {\color{black}III}, where the system was cooled from $545$ K down to $155$ K and back. On this part the SWCNTs show a completely metallic behavior (resistivity decreases upon decreasing temperature and increases upon increasing) proving that the intercalation took place. Further doping steps, {\color{OliveGreen}IV and VI} and measurement parts, {\color{black}V and VII} indicate the drop of the resistivity as the material is turning more and more metallic.}
\label{fig:k-swcnt}
\end{figure}

The next advantage of the microwave conductivity measurements is that they are readily adaptable to make \emph{in-situ} measurements. We demonstrate this on an arc-discharge SWCNT sample in the high temperature setup. The intercalation takes place similarly like in the two chamber vapor-phase intercalation method: the intercalant is separated in space and the process is driven by the gradient in temperature and chemical potential \cite{Dresselhaus1981}. The difference is that the nanotube powder is now placed in the middle of the cavity, while the intercalant, in our case the potassium, is located outside. The hot nitrogen melts the potassium and sustains the vapor pressure required for the doping. In this geometry the resistivity change of the material inside is measured instantly. In Figure \ref{fig:k-swcnt}. a complete \emph{in-situ} measurement process is presented.

The process is carried out in the following steps: first the temperature of the steadily cooled sample was increased from $173$ K to $520$ K, this part is labeled as {\color{black}I} in Fig. \ref{fig:k-swcnt}. In this range the resistivity of the material is decreasing upon increasing temperature, clearly showing a semiconducting behavior with $\rho = \rho_0 \expe^{\Delta/T}$. This is expected as the used arc-discharge SWCNT bundle contains a mixture of $2/3$ non-metallic and $1/3$ metallic nanotubes. A fitted exponential to this regime yields a transport activation energy of $\Delta = 39(1)$ K in agreement with previous observations \cite{KarsaPssb2012,KaiserPRB1998}. At $520$ K the intercalation starts to take place resulting in a decrease in the resistivity. The temperature is then fixed at a value of $545$ K for $16$ minutes, this regime is noted as {\color{OliveGreen}II} in the figure. Cooling down the sample ({\color{black}III}) we observe a completely different behavior as before: the sample becomes metallic, as its resistivity decreases with lowering the temperature. Afterwards the system is heated up again and the doping process is continued. The conductivity of the sample is increased further in {\color{OliveGreen}IV} as the sample gets more doped. In the further measurement and intercalation steps the resistivity drops further and further ({\color{black}V}, {\color{OliveGreen}VI} and {\color{black}VII}) and the process can be continued until reaching the equilibrium stoichiometry of KC$_7$, which is a completely metallic state \cite{Pichler1999}.

\section{Summary}

We presented that physical properties of air-sensitive nanomaterials, especially in powder form can be investigated in the framework of microwave cavity measurements, which appeared to be very robust and versatile. Conductivity measurements made on black phosphorus, KP$_8$, two types of K$_3$C$_{60}$ samples and real time evolution of K-SWCNTs, {\color{black}potassium intercalated single-walled nanotubes} are demonstrated herein.

\section{Acknowledgement}
Work supported by the Hungarian National Research, Development and Innovation Office (NKFIH) Grant Nr. K119442, SNN118012 and 2017-1.2.1-NKP-2017-00001. The authors thank the European Research Council (ERC Advanced Grant 742145 B-PhosphoChem) for financial support. The research leading to these results was partially funded by the European Union Seventh Framework Programme under grant agreement No. 604391 Graphene Flagship. We also thank the Deutsche Forschungsgemeinschaft (DFG-SFB 953 "Synthetic Carbon Allotropes", Project A1), the Interdisciplinary Center for Molecular Materials (ICMM), and the Graduate School Molecular Science (GSMS) for financial support. G. A. thanks the FAU for the Emerging Talents Initiative (ETI) grant \#WS16-17\_Nat\_04, and support by the DFG and FLAG-ERA (AB694/2-1).

\bibliography{mwcn}

\begin{thebibliography}{54}%
\makeatletter
\providecommand \@ifxundefined [1]{%
 \@ifx{#1\undefined}
}%
\providecommand \@ifnum [1]{%
 \ifnum #1\expandafter \@firstoftwo
 \else \expandafter \@secondoftwo
 \fi
}%
\providecommand \@ifx [1]{%
 \ifx #1\expandafter \@firstoftwo
 \else \expandafter \@secondoftwo
 \fi
}%
\providecommand \natexlab [1]{#1}%
\providecommand \enquote  [1]{``#1''}%
\providecommand \bibnamefont  [1]{#1}%
\providecommand \bibfnamefont [1]{#1}%
\providecommand \citenamefont [1]{#1}%
\providecommand \href@noop [0]{\@secondoftwo}%
\providecommand \href [0]{\begingroup \@sanitize@url \@href}%
\providecommand \@href[1]{\@@startlink{#1}\@@href}%
\providecommand \@@href[1]{\endgroup#1\@@endlink}%
\providecommand \@sanitize@url [0]{\catcode `\\12\catcode `\$12\catcode
  `\&12\catcode `\#12\catcode `\^12\catcode `\_12\catcode `\%12\relax}%
\providecommand \@@startlink[1]{}%
\providecommand \@@endlink[0]{}%
\providecommand \url  [0]{\begingroup\@sanitize@url \@url }%
\providecommand \@url [1]{\endgroup\@href {#1}{\urlprefix }}%
\providecommand \urlprefix  [0]{URL }%
\providecommand \Eprint [0]{\href }%
\providecommand \doibase [0]{http://dx.doi.org/}%
\providecommand \selectlanguage [0]{\@gobble}%
\providecommand \bibinfo  [0]{\@secondoftwo}%
\providecommand \bibfield  [0]{\@secondoftwo}%
\providecommand \translation [1]{[#1]}%
\providecommand \BibitemOpen [0]{}%
\providecommand \bibitemStop [0]{}%
\providecommand \bibitemNoStop [0]{.\EOS\space}%
\providecommand \EOS [0]{\spacefactor3000\relax}%
\providecommand \BibitemShut  [1]{\csname bibitem#1\endcsname}%
\let\auto@bib@innerbib\@empty
\bibitem [{\citenamefont {Kroto}\ \emph {et~al.}(1985)\citenamefont {Kroto},
  \citenamefont {Heath}, \citenamefont {O'Brien}, \citenamefont {Curl},\ and\
  \citenamefont {Smalley}}]{Kroto1985}%
  \BibitemOpen
  \bibfield  {author} {\bibinfo {author} {\bibfnamefont {H.~W.}\ \bibnamefont
  {Kroto}}, \bibinfo {author} {\bibfnamefont {J.~R.}\ \bibnamefont {Heath}},
  \bibinfo {author} {\bibfnamefont {S.~C.}\ \bibnamefont {O'Brien}}, \bibinfo
  {author} {\bibfnamefont {R.~F.}\ \bibnamefont {Curl}}, \ and\ \bibinfo
  {author} {\bibfnamefont {R.~E.}\ \bibnamefont {Smalley}},\ }\href@noop {}
  {\bibfield  {journal} {\bibinfo  {journal} {Nature}\ }\textbf {\bibinfo
  {volume} {318}},\ \bibinfo {pages} {162} (\bibinfo {year}
  {1985})}\BibitemShut {NoStop}%
\bibitem [{\citenamefont {Iijima}(1991)}]{Iijima1991}%
  \BibitemOpen
  \bibfield  {author} {\bibinfo {author} {\bibfnamefont {S.}~\bibnamefont
  {Iijima}},\ }\href@noop {} {\bibfield  {journal} {\bibinfo  {journal}
  {Nature}\ }\textbf {\bibinfo {volume} {354}},\ \bibinfo {pages} {56}
  (\bibinfo {year} {1991})}\BibitemShut {NoStop}%
\bibitem [{\citenamefont {Novoselov}\ \emph {et~al.}(2004)\citenamefont
  {Novoselov}, \citenamefont {Geim}, \citenamefont {Morozov}, \citenamefont
  {Jiang}, \citenamefont {Zhang}, \citenamefont {Dubonos}, \citenamefont
  {Grigorieva},\ and\ \citenamefont {Firsov}}]{Novoselov2004}%
  \BibitemOpen
  \bibfield  {author} {\bibinfo {author} {\bibfnamefont {K.~S.}\ \bibnamefont
  {Novoselov}}, \bibinfo {author} {\bibfnamefont {A.~K.}\ \bibnamefont {Geim}},
  \bibinfo {author} {\bibfnamefont {S.~V.}\ \bibnamefont {Morozov}}, \bibinfo
  {author} {\bibfnamefont {D.}~\bibnamefont {Jiang}}, \bibinfo {author}
  {\bibfnamefont {Y.}~\bibnamefont {Zhang}}, \bibinfo {author} {\bibfnamefont
  {S.~V.}\ \bibnamefont {Dubonos}}, \bibinfo {author} {\bibfnamefont {I.~V.}\
  \bibnamefont {Grigorieva}}, \ and\ \bibinfo {author} {\bibfnamefont {A.~A.}\
  \bibnamefont {Firsov}},\ }\href@noop {} {\bibfield  {journal} {\bibinfo
  {journal} {Science}\ }\textbf {\bibinfo {volume} {306}},\ \bibinfo {pages}
  {666} (\bibinfo {year} {2004})}\BibitemShut {NoStop}%
\bibitem [{\citenamefont {Li}\ \emph {et~al.}(2014)\citenamefont {Li},
  \citenamefont {Yu}, \citenamefont {Ye}, \citenamefont {Ge}, \citenamefont
  {Ou}, \citenamefont {Wu}, \citenamefont {Feng}, \citenamefont {Chen},\ and\
  \citenamefont {Zhang}}]{Li2014}%
  \BibitemOpen
  \bibfield  {author} {\bibinfo {author} {\bibfnamefont {L.}~\bibnamefont
  {Li}}, \bibinfo {author} {\bibfnamefont {Y.}~\bibnamefont {Yu}}, \bibinfo
  {author} {\bibfnamefont {G.~J.}\ \bibnamefont {Ye}}, \bibinfo {author}
  {\bibfnamefont {Q.}~\bibnamefont {Ge}}, \bibinfo {author} {\bibfnamefont
  {X.}~\bibnamefont {Ou}}, \bibinfo {author} {\bibfnamefont {H.}~\bibnamefont
  {Wu}}, \bibinfo {author} {\bibfnamefont {D.}~\bibnamefont {Feng}}, \bibinfo
  {author} {\bibfnamefont {X.~H.}\ \bibnamefont {Chen}}, \ and\ \bibinfo
  {author} {\bibfnamefont {Y.}~\bibnamefont {Zhang}},\ }\href@noop {}
  {\bibfield  {journal} {\bibinfo  {journal} {Nature Nanotechnology}\ }\textbf
  {\bibinfo {volume} {9}},\ \bibinfo {pages} {372} (\bibinfo {year}
  {2014})}\BibitemShut {NoStop}%
\bibitem [{\citenamefont {Liu}\ \emph {et~al.}(2014)\citenamefont {Liu},
  \citenamefont {Neal}, \citenamefont {Zhu}, \citenamefont {Luo}, \citenamefont
  {Xu}, \citenamefont {Tom\'{a}nek},\ and\ \citenamefont {Ye}}]{Liu2014}%
  \BibitemOpen
  \bibfield  {author} {\bibinfo {author} {\bibfnamefont {H.}~\bibnamefont
  {Liu}}, \bibinfo {author} {\bibfnamefont {A.~T.}\ \bibnamefont {Neal}},
  \bibinfo {author} {\bibfnamefont {Z.}~\bibnamefont {Zhu}}, \bibinfo {author}
  {\bibfnamefont {Z.}~\bibnamefont {Luo}}, \bibinfo {author} {\bibfnamefont
  {X.}~\bibnamefont {Xu}}, \bibinfo {author} {\bibfnamefont {D.}~\bibnamefont
  {Tom\'{a}nek}}, \ and\ \bibinfo {author} {\bibfnamefont {P.~D.}\ \bibnamefont
  {Ye}},\ }\href@noop {} {\bibfield  {journal} {\bibinfo  {journal} {ACS Nano}\
  }\textbf {\bibinfo {volume} {8}},\ \bibinfo {pages} {4033} (\bibinfo {year}
  {2014})}\BibitemShut {NoStop}%
\bibitem [{\citenamefont {Castellanos-Gomez}(2015)}]{castellanos-gomez2015}%
  \BibitemOpen
  \bibfield  {author} {\bibinfo {author} {\bibfnamefont {A.}~\bibnamefont
  {Castellanos-Gomez}},\ }\href@noop {} {\bibfield  {journal} {\bibinfo
  {journal} {The Journal of Physical Chemistry Letters}\ }\textbf {\bibinfo
  {volume} {6}},\ \bibinfo {pages} {4280} (\bibinfo {year} {2015})}\BibitemShut
  {NoStop}%
\bibitem [{\citenamefont {Ling}\ \emph {et~al.}(2015)\citenamefont {Ling},
  \citenamefont {Wang}, \citenamefont {Huang}, \citenamefont {Xia},\ and\
  \citenamefont {Dresselhaus}}]{ling2015}%
  \BibitemOpen
  \bibfield  {author} {\bibinfo {author} {\bibfnamefont {X.}~\bibnamefont
  {Ling}}, \bibinfo {author} {\bibfnamefont {H.}~\bibnamefont {Wang}}, \bibinfo
  {author} {\bibfnamefont {S.}~\bibnamefont {Huang}}, \bibinfo {author}
  {\bibfnamefont {F.}~\bibnamefont {Xia}}, \ and\ \bibinfo {author}
  {\bibfnamefont {M.~S.}\ \bibnamefont {Dresselhaus}},\ }\href@noop {}
  {\bibfield  {journal} {\bibinfo  {journal} {Proceedings of the National
  Academy of Sciences of the United States of America (PNAS)}\ }\textbf
  {\bibinfo {volume} {112}},\ \bibinfo {pages} {4523} (\bibinfo {year}
  {2015})}\BibitemShut {NoStop}%
\bibitem [{\citenamefont {Liu}\ \emph {et~al.}(2015)\citenamefont {Liu},
  \citenamefont {Du}, \citenamefont {Deng},\ and\ \citenamefont
  {Ye}}]{Liu2015}%
  \BibitemOpen
  \bibfield  {author} {\bibinfo {author} {\bibfnamefont {H.}~\bibnamefont
  {Liu}}, \bibinfo {author} {\bibfnamefont {Y.}~\bibnamefont {Du}}, \bibinfo
  {author} {\bibfnamefont {Y.}~\bibnamefont {Deng}}, \ and\ \bibinfo {author}
  {\bibfnamefont {P.~D.}\ \bibnamefont {Ye}},\ }\href@noop {} {\bibfield
  {journal} {\bibinfo  {journal} {Chemical Society Reviews}\ }\textbf {\bibinfo
  {volume} {44}},\ \bibinfo {pages} {2732} (\bibinfo {year}
  {2015})}\BibitemShut {NoStop}%
\bibitem [{\citenamefont {Ryder}\ \emph {et~al.}(2016)\citenamefont {Ryder},
  \citenamefont {Wood}, \citenamefont {Wells},\ and\ \citenamefont
  {Hersam}}]{Ryder2016}%
  \BibitemOpen
  \bibfield  {author} {\bibinfo {author} {\bibfnamefont {C.~R.}\ \bibnamefont
  {Ryder}}, \bibinfo {author} {\bibfnamefont {J.~D.}\ \bibnamefont {Wood}},
  \bibinfo {author} {\bibfnamefont {S.~A.}\ \bibnamefont {Wells}}, \ and\
  \bibinfo {author} {\bibfnamefont {M.~C.}\ \bibnamefont {Hersam}},\
  }\href@noop {} {\bibfield  {journal} {\bibinfo  {journal} {ACS Nano}\
  }\textbf {\bibinfo {volume} {10}},\ \bibinfo {pages} {3900} (\bibinfo {year}
  {2016})}\BibitemShut {NoStop}%
\bibitem [{\citenamefont {Gusm$\tilde{\text{a}}$o}\ \emph
  {et~al.}(2017)\citenamefont {Gusm$\tilde{\text{a}}$o}, \citenamefont
  {Sofer},\ and\ \citenamefont {Pumera}}]{gusmao2017}%
  \BibitemOpen
  \bibfield  {author} {\bibinfo {author} {\bibfnamefont {R.}~\bibnamefont
  {Gusm$\tilde{\text{a}}$o}}, \bibinfo {author} {\bibfnamefont
  {Z.}~\bibnamefont {Sofer}}, \ and\ \bibinfo {author} {\bibfnamefont
  {M.}~\bibnamefont {Pumera}},\ }\href@noop {} {\bibfield  {journal} {\bibinfo
  {journal} {Angewande Chemie International Edition}\ }\textbf {\bibinfo
  {volume} {56}},\ \bibinfo {pages} {8052} (\bibinfo {year}
  {2017})}\BibitemShut {NoStop}%
\bibitem [{\citenamefont {Hebard}\ \emph {et~al.}(1991)\citenamefont {Hebard},
  \citenamefont {Rosseinsky}, \citenamefont {Haddon}, \citenamefont {Murphy},
  \citenamefont {Glarum}, \citenamefont {Palstra}, \citenamefont {Ramirez},\
  and\ \citenamefont {Kortan}}]{Hebard1991}%
  \BibitemOpen
  \bibfield  {author} {\bibinfo {author} {\bibfnamefont {A.~F.}\ \bibnamefont
  {Hebard}}, \bibinfo {author} {\bibfnamefont {M.~J.}\ \bibnamefont
  {Rosseinsky}}, \bibinfo {author} {\bibfnamefont {R.~C.}\ \bibnamefont
  {Haddon}}, \bibinfo {author} {\bibfnamefont {D.~W.}\ \bibnamefont {Murphy}},
  \bibinfo {author} {\bibfnamefont {S.~H.}\ \bibnamefont {Glarum}}, \bibinfo
  {author} {\bibfnamefont {T.~T.~M.}\ \bibnamefont {Palstra}}, \bibinfo
  {author} {\bibfnamefont {A.~P.}\ \bibnamefont {Ramirez}}, \ and\ \bibinfo
  {author} {\bibfnamefont {A.~R.}\ \bibnamefont {Kortan}},\ }\href@noop {}
  {\bibfield  {journal} {\bibinfo  {journal} {Nature}\ }\textbf {\bibinfo
  {volume} {350}},\ \bibinfo {pages} {600} (\bibinfo {year}
  {1991})}\BibitemShut {NoStop}%
\bibitem [{\citenamefont {Bommeli}\ \emph {et~al.}(1995)\citenamefont
  {Bommeli}, \citenamefont {Degiorgi}, \citenamefont {Wachter}, \citenamefont
  {Legeza}, \citenamefont {J\'anossy}, \citenamefont {Oszlanyi}, \citenamefont
  {Chauvet},\ and\ \citenamefont {Forro}}]{Bommeli1995}%
  \BibitemOpen
  \bibfield  {author} {\bibinfo {author} {\bibfnamefont {F.}~\bibnamefont
  {Bommeli}}, \bibinfo {author} {\bibfnamefont {L.}~\bibnamefont {Degiorgi}},
  \bibinfo {author} {\bibfnamefont {P.}~\bibnamefont {Wachter}}, \bibinfo
  {author} {\bibfnamefont {O.}~\bibnamefont {Legeza}}, \bibinfo {author}
  {\bibfnamefont {A.}~\bibnamefont {J\'anossy}}, \bibinfo {author}
  {\bibfnamefont {G.}~\bibnamefont {Oszlanyi}}, \bibinfo {author}
  {\bibfnamefont {O.}~\bibnamefont {Chauvet}}, \ and\ \bibinfo {author}
  {\bibfnamefont {L.}~\bibnamefont {Forro}},\ }\href@noop {} {\bibfield
  {journal} {\bibinfo  {journal} {Physical Review B}\ }\textbf {\bibinfo
  {volume} {51}},\ \bibinfo {pages} {14794} (\bibinfo {year}
  {1995})}\BibitemShut {NoStop}%
\bibitem [{\citenamefont {Kitano}\ \emph
  {et~al.}(2002{\natexlab{a}})\citenamefont {Kitano}, \citenamefont {Matsuo},
  \citenamefont {Miwa}, \citenamefont {Maeda}, \citenamefont {Takenobu},
  \citenamefont {Iwasa},\ and\ \citenamefont {Mitani}}]{Maeda2002}%
  \BibitemOpen
  \bibfield  {author} {\bibinfo {author} {\bibfnamefont {H.}~\bibnamefont
  {Kitano}}, \bibinfo {author} {\bibfnamefont {R.}~\bibnamefont {Matsuo}},
  \bibinfo {author} {\bibfnamefont {K.}~\bibnamefont {Miwa}}, \bibinfo {author}
  {\bibfnamefont {A.}~\bibnamefont {Maeda}}, \bibinfo {author} {\bibfnamefont
  {T.}~\bibnamefont {Takenobu}}, \bibinfo {author} {\bibfnamefont
  {Y.}~\bibnamefont {Iwasa}}, \ and\ \bibinfo {author} {\bibfnamefont
  {T.}~\bibnamefont {Mitani}},\ }\href@noop {} {\bibfield  {journal} {\bibinfo
  {journal} {Physical Review Letters}\ }\textbf {\bibinfo {volume} {88}},\
  \bibinfo {pages} {096401} (\bibinfo {year} {2002}{\natexlab{a}})}\BibitemShut
  {NoStop}%
\bibitem [{\citenamefont {Rao}\ \emph {et~al.}(1997)\citenamefont {Rao},
  \citenamefont {Eklund}, \citenamefont {Bandow}, \citenamefont {Thess},\ and\
  \citenamefont {Smalley}}]{Eklund1997}%
  \BibitemOpen
  \bibfield  {author} {\bibinfo {author} {\bibfnamefont {A.~M.}\ \bibnamefont
  {Rao}}, \bibinfo {author} {\bibfnamefont {P.~C.}\ \bibnamefont {Eklund}},
  \bibinfo {author} {\bibfnamefont {S.}~\bibnamefont {Bandow}}, \bibinfo
  {author} {\bibfnamefont {A.}~\bibnamefont {Thess}}, \ and\ \bibinfo {author}
  {\bibfnamefont {R.~E.}\ \bibnamefont {Smalley}},\ }\href@noop {} {\bibfield
  {journal} {\bibinfo  {journal} {Nature}\ }\textbf {\bibinfo {volume} {388}},\
  \bibinfo {pages} {257} (\bibinfo {year} {1997})}\BibitemShut {NoStop}%
\bibitem [{\citenamefont {Rauf}\ \emph {et~al.}(2004)\citenamefont {Rauf},
  \citenamefont {Pichler}, \citenamefont {Knupfer}, \citenamefont {Fink},\ and\
  \citenamefont {Kataura}}]{Pichler2004}%
  \BibitemOpen
  \bibfield  {author} {\bibinfo {author} {\bibfnamefont {H.}~\bibnamefont
  {Rauf}}, \bibinfo {author} {\bibfnamefont {T.}~\bibnamefont {Pichler}},
  \bibinfo {author} {\bibfnamefont {M.}~\bibnamefont {Knupfer}}, \bibinfo
  {author} {\bibfnamefont {J.}~\bibnamefont {Fink}}, \ and\ \bibinfo {author}
  {\bibfnamefont {H.}~\bibnamefont {Kataura}},\ }\href@noop {} {\bibfield
  {journal} {\bibinfo  {journal} {Physical Review Letters}\ }\textbf {\bibinfo
  {volume} {93}},\ \bibinfo {pages} {096805} (\bibinfo {year}
  {2004})}\BibitemShut {NoStop}%
\bibitem [{\citenamefont {Ziletti}\ \emph {et~al.}(2015)\citenamefont
  {Ziletti}, \citenamefont {Carvalho}, \citenamefont {Campbell}, \citenamefont
  {Coker},\ and\ \citenamefont {{{Castro~Neto}}}}]{ziletti2015}%
  \BibitemOpen
  \bibfield  {author} {\bibinfo {author} {\bibfnamefont {A.}~\bibnamefont
  {Ziletti}}, \bibinfo {author} {\bibfnamefont {A.}~\bibnamefont {Carvalho}},
  \bibinfo {author} {\bibfnamefont {D.~K.}\ \bibnamefont {Campbell}}, \bibinfo
  {author} {\bibfnamefont {D.~F.}\ \bibnamefont {Coker}}, \ and\ \bibinfo
  {author} {\bibfnamefont {A.~H.}\ \bibnamefont {{{Castro~Neto}}}},\
  }\href@noop {} {\bibfield  {journal} {\bibinfo  {journal} {Physical Review
  Letters}\ }\textbf {\bibinfo {volume} {114}},\ \bibinfo {pages} {046801}
  (\bibinfo {year} {2015})}\BibitemShut {NoStop}%
\bibitem [{\citenamefont {Abell\'an}\ \emph
  {et~al.}(2017{\natexlab{a}})\citenamefont {Abell\'an}, \citenamefont {Wild},
  \citenamefont {Lloret}, \citenamefont {Scheuschner}, \citenamefont {Gillen},
  \citenamefont {Mundloch}, \citenamefont {Maultzsch}, \citenamefont {Varela},
  \citenamefont {Hauke},\ and\ \citenamefont {Hirsch}}]{AbellanJACS2017}%
  \BibitemOpen
  \bibfield  {author} {\bibinfo {author} {\bibfnamefont {G.}~\bibnamefont
  {Abell\'an}}, \bibinfo {author} {\bibfnamefont {S.}~\bibnamefont {Wild}},
  \bibinfo {author} {\bibfnamefont {V.}~\bibnamefont {Lloret}}, \bibinfo
  {author} {\bibfnamefont {N.}~\bibnamefont {Scheuschner}}, \bibinfo {author}
  {\bibfnamefont {R.}~\bibnamefont {Gillen}}, \bibinfo {author} {\bibfnamefont
  {U.}~\bibnamefont {Mundloch}}, \bibinfo {author} {\bibfnamefont
  {J.}~\bibnamefont {Maultzsch}}, \bibinfo {author} {\bibfnamefont
  {M.}~\bibnamefont {Varela}}, \bibinfo {author} {\bibfnamefont
  {F.}~\bibnamefont {Hauke}}, \ and\ \bibinfo {author} {\bibfnamefont
  {A.}~\bibnamefont {Hirsch}},\ }\href@noop {} {\bibfield  {journal} {\bibinfo
  {journal} {Journal of the Americal Chemical Society}\ }\textbf {\bibinfo
  {volume} {139}},\ \bibinfo {pages} {10432} (\bibinfo {year}
  {2017}{\natexlab{a}})}\BibitemShut {NoStop}%
\bibitem [{\citenamefont {Zhang}\ \emph {et~al.}(2018)\citenamefont {Zhang},
  \citenamefont {Wan}, \citenamefont {Xie}, \citenamefont {Mu}, \citenamefont
  {Du}, \citenamefont {Wang}, \citenamefont {Wu}, \citenamefont {Ji},\ and\
  \citenamefont {Wan}}]{ZhangJACS2018}%
  \BibitemOpen
  \bibfield  {author} {\bibinfo {author} {\bibfnamefont {T.}~\bibnamefont
  {Zhang}}, \bibinfo {author} {\bibfnamefont {Y.}~\bibnamefont {Wan}}, \bibinfo
  {author} {\bibfnamefont {H.}~\bibnamefont {Xie}}, \bibinfo {author}
  {\bibfnamefont {Y.}~\bibnamefont {Mu}}, \bibinfo {author} {\bibfnamefont
  {P.}~\bibnamefont {Du}}, \bibinfo {author} {\bibfnamefont {D.}~\bibnamefont
  {Wang}}, \bibinfo {author} {\bibfnamefont {X.}~\bibnamefont {Wu}}, \bibinfo
  {author} {\bibfnamefont {H.}~\bibnamefont {Ji}}, \ and\ \bibinfo {author}
  {\bibfnamefont {L.}~\bibnamefont {Wan}},\ }\href@noop {} {\bibfield
  {journal} {\bibinfo  {journal} {Journal of the Americal Chemical Society}\
  }\textbf {\bibinfo {volume} {jacs.8b02156}} (\bibinfo {year}
  {2018})}\BibitemShut {NoStop}%
\bibitem [{\citenamefont {Scheffler}\ \emph {et~al.}(2005)\citenamefont
  {Scheffler}, \citenamefont {Dressel}, \citenamefont {Jourdan},\ and\
  \citenamefont {Adrian}}]{Dressel2005}%
  \BibitemOpen
  \bibfield  {author} {\bibinfo {author} {\bibfnamefont {M.}~\bibnamefont
  {Scheffler}}, \bibinfo {author} {\bibfnamefont {M.}~\bibnamefont {Dressel}},
  \bibinfo {author} {\bibfnamefont {M.}~\bibnamefont {Jourdan}}, \ and\
  \bibinfo {author} {\bibfnamefont {H.}~\bibnamefont {Adrian}},\ }\href@noop {}
  {\bibfield  {journal} {\bibinfo  {journal} {Nature}\ }\textbf {\bibinfo
  {volume} {438}},\ \bibinfo {pages} {1135} (\bibinfo {year}
  {2005})}\BibitemShut {NoStop}%
\bibitem [{\citenamefont {Scheffler}\ \emph {et~al.}(2013)\citenamefont
  {Scheffler}, \citenamefont {Schlegel}, \citenamefont {Clauss}, \citenamefont
  {Hafner}, \citenamefont {Fella}, \citenamefont {Dressel}, \citenamefont
  {Jourdan}, \citenamefont {Sichelschmidt}, \citenamefont {Krellner},
  \citenamefont {Geibel},\ and\ \citenamefont {Steglich}}]{Dressel2013}%
  \BibitemOpen
  \bibfield  {author} {\bibinfo {author} {\bibfnamefont {M.}~\bibnamefont
  {Scheffler}}, \bibinfo {author} {\bibfnamefont {K.}~\bibnamefont {Schlegel}},
  \bibinfo {author} {\bibfnamefont {C.}~\bibnamefont {Clauss}}, \bibinfo
  {author} {\bibfnamefont {D.}~\bibnamefont {Hafner}}, \bibinfo {author}
  {\bibfnamefont {C.}~\bibnamefont {Fella}}, \bibinfo {author} {\bibfnamefont
  {M.}~\bibnamefont {Dressel}}, \bibinfo {author} {\bibfnamefont
  {M.}~\bibnamefont {Jourdan}}, \bibinfo {author} {\bibfnamefont
  {J.}~\bibnamefont {Sichelschmidt}}, \bibinfo {author} {\bibfnamefont
  {C.}~\bibnamefont {Krellner}}, \bibinfo {author} {\bibfnamefont
  {C.}~\bibnamefont {Geibel}}, \ and\ \bibinfo {author} {\bibfnamefont
  {F.}~\bibnamefont {Steglich}},\ }\href@noop {} {\bibfield  {journal}
  {\bibinfo  {journal} {physica status solidi (b)}\ }\textbf {\bibinfo {volume}
  {250}},\ \bibinfo {pages} {439} (\bibinfo {year} {2013})}\BibitemShut
  {NoStop}%
\bibitem [{\citenamefont {Buravov}\ and\ \citenamefont
  {Shchegolev}(1971)}]{Buravov1971}%
  \BibitemOpen
  \bibfield  {author} {\bibinfo {author} {\bibfnamefont {L.~I.}\ \bibnamefont
  {Buravov}}\ and\ \bibinfo {author} {\bibfnamefont {I.~F.}\ \bibnamefont
  {Shchegolev}},\ }\href@noop {} {\bibfield  {journal} {\bibinfo  {journal}
  {Instruments and Experimental Techniques}\ }\textbf {\bibinfo {volume}
  {14}},\ \bibinfo {pages} {528} (\bibinfo {year} {1971})}\BibitemShut
  {NoStop}%
\bibitem [{\citenamefont {Klein}\ \emph {et~al.}(1993)\citenamefont {Klein},
  \citenamefont {Donovan}, \citenamefont {Dressel},\ and\ \citenamefont
  {Gr{\"u}ner}}]{Klein1993}%
  \BibitemOpen
  \bibfield  {author} {\bibinfo {author} {\bibfnamefont {O.}~\bibnamefont
  {Klein}}, \bibinfo {author} {\bibfnamefont {S.}~\bibnamefont {Donovan}},
  \bibinfo {author} {\bibfnamefont {M.}~\bibnamefont {Dressel}}, \ and\
  \bibinfo {author} {\bibfnamefont {G.}~\bibnamefont {Gr{\"u}ner}},\
  }\href@noop {} {\bibfield  {journal} {\bibinfo  {journal} {International
  Journal of Infrared and Millimeter Waves}\ }\textbf {\bibinfo {volume}
  {14}},\ \bibinfo {pages} {2423} (\bibinfo {year} {1993})}\BibitemShut
  {NoStop}%
\bibitem [{\citenamefont {Klein}\ \emph
  {et~al.}(1994{\natexlab{a}})\citenamefont {Klein}, \citenamefont {Nicol},
  \citenamefont {Holczer},\ and\ \citenamefont {Gr\"uner}}]{HolczerPRB}%
  \BibitemOpen
  \bibfield  {author} {\bibinfo {author} {\bibfnamefont {O.}~\bibnamefont
  {Klein}}, \bibinfo {author} {\bibfnamefont {E.~J.}\ \bibnamefont {Nicol}},
  \bibinfo {author} {\bibfnamefont {K.}~\bibnamefont {Holczer}}, \ and\
  \bibinfo {author} {\bibfnamefont {G.}~\bibnamefont {Gr\"uner}},\ }\href@noop
  {} {\bibfield  {journal} {\bibinfo  {journal} {Physical Review B}\ }\textbf
  {\bibinfo {volume} {50}},\ \bibinfo {pages} {6307} (\bibinfo {year}
  {1994}{\natexlab{a}})}\BibitemShut {NoStop}%
\bibitem [{\citenamefont {Kitano}\ \emph
  {et~al.}(2002{\natexlab{b}})\citenamefont {Kitano}, \citenamefont {Matsuo},
  \citenamefont {Miwa}, \citenamefont {Maeda}, \citenamefont {Takenobu},
  \citenamefont {Iwasa},\ and\ \citenamefont {Mitani}}]{MaedaPRL}%
  \BibitemOpen
  \bibfield  {author} {\bibinfo {author} {\bibfnamefont {H.}~\bibnamefont
  {Kitano}}, \bibinfo {author} {\bibfnamefont {R.}~\bibnamefont {Matsuo}},
  \bibinfo {author} {\bibfnamefont {K.}~\bibnamefont {Miwa}}, \bibinfo {author}
  {\bibfnamefont {A.}~\bibnamefont {Maeda}}, \bibinfo {author} {\bibfnamefont
  {T.}~\bibnamefont {Takenobu}}, \bibinfo {author} {\bibfnamefont
  {Y.}~\bibnamefont {Iwasa}}, \ and\ \bibinfo {author} {\bibfnamefont
  {T.}~\bibnamefont {Mitani}},\ }\href@noop {} {\bibfield  {journal} {\bibinfo
  {journal} {Physical Review Letters}\ }\textbf {\bibinfo {volume} {88}},\
  \bibinfo {pages} {096401} (\bibinfo {year} {2002}{\natexlab{b}})}\BibitemShut
  {NoStop}%
\bibitem [{\citenamefont {Dressel}(2013)}]{DresselReview}%
  \BibitemOpen
  \bibfield  {author} {\bibinfo {author} {\bibfnamefont {M.}~\bibnamefont
  {Dressel}},\ }\href@noop {} {\bibfield  {journal} {\bibinfo  {journal}
  {Advances in Condensed Matter Physics}\ } (\bibinfo {year}
  {2013})}\BibitemShut {NoStop}%
\bibitem [{\citenamefont {Abell\'an}\ \emph
  {et~al.}(2017{\natexlab{b}})\citenamefont {Abell\'an}, \citenamefont {Neiss},
  \citenamefont {Lloret}, \citenamefont {Wild}, \citenamefont
  {Chac\'on‐Torres}, \citenamefont {Werbach}, \citenamefont {Fedi},
  \citenamefont {Shiozawa}, \citenamefont {G\"orling}, \citenamefont
  {Peterlik}, \citenamefont {Pichler}, \citenamefont {Hauke},\ and\
  \citenamefont {Hirsch}}]{Abellan2017}%
  \BibitemOpen
  \bibfield  {author} {\bibinfo {author} {\bibfnamefont {G.}~\bibnamefont
  {Abell\'an}}, \bibinfo {author} {\bibfnamefont {C.}~\bibnamefont {Neiss}},
  \bibinfo {author} {\bibfnamefont {V.}~\bibnamefont {Lloret}}, \bibinfo
  {author} {\bibfnamefont {S.}~\bibnamefont {Wild}}, \bibinfo {author}
  {\bibfnamefont {J.~C.}\ \bibnamefont {Chac\'on‐Torres}}, \bibinfo {author}
  {\bibfnamefont {K.}~\bibnamefont {Werbach}}, \bibinfo {author} {\bibfnamefont
  {F.}~\bibnamefont {Fedi}}, \bibinfo {author} {\bibfnamefont {H.}~\bibnamefont
  {Shiozawa}}, \bibinfo {author} {\bibfnamefont {A.}~\bibnamefont {G\"orling}},
  \bibinfo {author} {\bibfnamefont {H.}~\bibnamefont {Peterlik}}, \bibinfo
  {author} {\bibfnamefont {T.}~\bibnamefont {Pichler}}, \bibinfo {author}
  {\bibfnamefont {F.}~\bibnamefont {Hauke}}, \ and\ \bibinfo {author}
  {\bibfnamefont {A.}~\bibnamefont {Hirsch}},\ }\href@noop {} {\bibfield
  {journal} {\bibinfo  {journal} {Angewandte Chemie International Edition}\
  }\textbf {\bibinfo {volume} {56}},\ \bibinfo {pages} {15267} (\bibinfo {year}
  {2017}{\natexlab{b}})}\BibitemShut {NoStop}%
\bibitem [{\citenamefont {Nemes}\ \emph {et~al.}(2000)\citenamefont {Nemes},
  \citenamefont {Fischer}, \citenamefont {Baumgartner}, \citenamefont
  {Forr\'o}, \citenamefont {Feh\'er}, \citenamefont {Oszl\'anyi}, \citenamefont
  {Simon},\ and\ \citenamefont {J\'anossy}}]{NemesPRB2000}%
  \BibitemOpen
  \bibfield  {author} {\bibinfo {author} {\bibfnamefont {N.~M.}\ \bibnamefont
  {Nemes}}, \bibinfo {author} {\bibfnamefont {J.~E.}\ \bibnamefont {Fischer}},
  \bibinfo {author} {\bibfnamefont {G.}~\bibnamefont {Baumgartner}}, \bibinfo
  {author} {\bibfnamefont {L.}~\bibnamefont {Forr\'o}}, \bibinfo {author}
  {\bibfnamefont {T.}~\bibnamefont {Feh\'er}}, \bibinfo {author} {\bibfnamefont
  {G.}~\bibnamefont {Oszl\'anyi}}, \bibinfo {author} {\bibfnamefont
  {F.}~\bibnamefont {Simon}}, \ and\ \bibinfo {author} {\bibfnamefont
  {A.}~\bibnamefont {J\'anossy}},\ }\href@noop {} {\bibfield  {journal}
  {\bibinfo  {journal} {Physical Review B}\ }\textbf {\bibinfo {volume} {61}},\
  \bibinfo {pages} {7118} (\bibinfo {year} {2000})}\BibitemShut {NoStop}%
\bibitem [{\citenamefont {Simon}\ \emph
  {et~al.}(2005{\natexlab{a}})\citenamefont {Simon}, \citenamefont {Kukovecz},
  \citenamefont {Kramberger}, \citenamefont {Pfeiffer}, \citenamefont {Hasi},
  \citenamefont {Kuzmany},\ and\ \citenamefont {Kataura}}]{SimonPRB2005}%
  \BibitemOpen
  \bibfield  {author} {\bibinfo {author} {\bibfnamefont {F.}~\bibnamefont
  {Simon}}, \bibinfo {author} {\bibfnamefont {{\'{A}}.}~\bibnamefont
  {Kukovecz}}, \bibinfo {author} {\bibfnamefont {C.}~\bibnamefont
  {Kramberger}}, \bibinfo {author} {\bibfnamefont {R.}~\bibnamefont
  {Pfeiffer}}, \bibinfo {author} {\bibfnamefont {F.}~\bibnamefont {Hasi}},
  \bibinfo {author} {\bibfnamefont {H.}~\bibnamefont {Kuzmany}}, \ and\
  \bibinfo {author} {\bibfnamefont {H.}~\bibnamefont {Kataura}},\ }\href@noop
  {} {\bibfield  {journal} {\bibinfo  {journal} {Physical Review B}\ }\textbf
  {\bibinfo {volume} {71}},\ \bibinfo {pages} {165439} (\bibinfo {year}
  {2005}{\natexlab{a}})}\BibitemShut {NoStop}%
\bibitem [{\citenamefont {Simon}\ \emph
  {et~al.}(2005{\natexlab{b}})\citenamefont {Simon}, \citenamefont
  {Kramberger}, \citenamefont {Pfeiffer}, \citenamefont {Kuzmany},
  \citenamefont {Z\'{o}lyomi}, \citenamefont {K\"{u}rti}, \citenamefont
  {Singer},\ and\ \citenamefont {Alloul}}]{SimonPRL2005}%
  \BibitemOpen
  \bibfield  {author} {\bibinfo {author} {\bibfnamefont {F.}~\bibnamefont
  {Simon}}, \bibinfo {author} {\bibfnamefont {C.}~\bibnamefont {Kramberger}},
  \bibinfo {author} {\bibfnamefont {R.}~\bibnamefont {Pfeiffer}}, \bibinfo
  {author} {\bibfnamefont {H.}~\bibnamefont {Kuzmany}}, \bibinfo {author}
  {\bibfnamefont {V.}~\bibnamefont {Z\'{o}lyomi}}, \bibinfo {author}
  {\bibfnamefont {J.}~\bibnamefont {K\"{u}rti}}, \bibinfo {author}
  {\bibfnamefont {P.~M.}\ \bibnamefont {Singer}}, \ and\ \bibinfo {author}
  {\bibfnamefont {H.}~\bibnamefont {Alloul}},\ }\href@noop {} {\bibfield
  {journal} {\bibinfo  {journal} {Physical Review Letters}\ }\textbf {\bibinfo
  {volume} {95}},\ \bibinfo {pages} {017401} (\bibinfo {year}
  {2005}{\natexlab{b}})}\BibitemShut {NoStop}%
\bibitem [{\citenamefont {Szirmai}\ \emph {et~al.}(2017)\citenamefont
  {Szirmai}, \citenamefont {M\'arkus}, \citenamefont {D\'ora}, \citenamefont
  {F\'abi\'an}, \citenamefont {Koltai}, \citenamefont {Z\'olyomi},
  \citenamefont {K\"urti}, \citenamefont {N\'afr\'adi}, \citenamefont
  {Forr\'o}, \citenamefont {Pichler},\ and\ \citenamefont
  {Simon}}]{SzirmaiPRB2017}%
  \BibitemOpen
  \bibfield  {author} {\bibinfo {author} {\bibfnamefont {P.}~\bibnamefont
  {Szirmai}}, \bibinfo {author} {\bibfnamefont {B.~G.}\ \bibnamefont
  {M\'arkus}}, \bibinfo {author} {\bibfnamefont {B.}~\bibnamefont {D\'ora}},
  \bibinfo {author} {\bibfnamefont {G.}~\bibnamefont {F\'abi\'an}}, \bibinfo
  {author} {\bibfnamefont {J.}~\bibnamefont {Koltai}}, \bibinfo {author}
  {\bibfnamefont {V.}~\bibnamefont {Z\'olyomi}}, \bibinfo {author}
  {\bibfnamefont {J.}~\bibnamefont {K\"urti}}, \bibinfo {author} {\bibfnamefont
  {B.}~\bibnamefont {N\'afr\'adi}}, \bibinfo {author} {\bibfnamefont
  {L.}~\bibnamefont {Forr\'o}}, \bibinfo {author} {\bibfnamefont
  {T.}~\bibnamefont {Pichler}}, \ and\ \bibinfo {author} {\bibfnamefont
  {F.}~\bibnamefont {Simon}},\ }\href@noop {} {\bibfield  {journal} {\bibinfo
  {journal} {Physical Review B}\ }\textbf {\bibinfo {volume} {96}},\ \bibinfo
  {pages} {075133} (\bibinfo {year} {2017})}\BibitemShut {NoStop}%
\bibitem [{\citenamefont {Simon}\ \emph {et~al.}(2006)\citenamefont {Simon},
  \citenamefont {Kuzmany}, \citenamefont {N\'{a}fr\'{a}di}, \citenamefont
  {Feh\'{e}r}, \citenamefont {Forr\'{o}}, \citenamefont {F\"{u}l\"{o}p},
  \citenamefont {J\'{a}nossy}, \citenamefont {Korecz}, \citenamefont
  {Rockenbauer}, \citenamefont {Hauke},\ and\ \citenamefont
  {Hirsch}}]{SimonPRL2006}%
  \BibitemOpen
  \bibfield  {author} {\bibinfo {author} {\bibfnamefont {F.}~\bibnamefont
  {Simon}}, \bibinfo {author} {\bibfnamefont {H.}~\bibnamefont {Kuzmany}},
  \bibinfo {author} {\bibfnamefont {B.}~\bibnamefont {N\'{a}fr\'{a}di}},
  \bibinfo {author} {\bibfnamefont {T.}~\bibnamefont {Feh\'{e}r}}, \bibinfo
  {author} {\bibfnamefont {L.}~\bibnamefont {Forr\'{o}}}, \bibinfo {author}
  {\bibfnamefont {F.}~\bibnamefont {F\"{u}l\"{o}p}}, \bibinfo {author}
  {\bibfnamefont {A.}~\bibnamefont {J\'{a}nossy}}, \bibinfo {author}
  {\bibfnamefont {L.}~\bibnamefont {Korecz}}, \bibinfo {author} {\bibfnamefont
  {A.}~\bibnamefont {Rockenbauer}}, \bibinfo {author} {\bibfnamefont
  {F.}~\bibnamefont {Hauke}}, \ and\ \bibinfo {author} {\bibfnamefont
  {A.}~\bibnamefont {Hirsch}},\ }\href@noop {} {\bibfield  {journal} {\bibinfo
  {journal} {Physical Review Letters}\ }\textbf {\bibinfo {volume} {97}},\
  \bibinfo {pages} {136801} (\bibinfo {year} {2006})}\BibitemShut {NoStop}%
\bibitem [{\citenamefont {Simon}\ \emph {et~al.}(2004)\citenamefont {Simon},
  \citenamefont {Kuzmany}, \citenamefont {Rauf}, \citenamefont {Pichler},
  \citenamefont {Bernardi}, \citenamefont {Peterlik}, \citenamefont {Korecz},
  \citenamefont {F\"ul\"op},\ and\ \citenamefont
  {J\'anossy}}]{SimonChemPhysLett2004}%
  \BibitemOpen
  \bibfield  {author} {\bibinfo {author} {\bibfnamefont {F.}~\bibnamefont
  {Simon}}, \bibinfo {author} {\bibfnamefont {H.}~\bibnamefont {Kuzmany}},
  \bibinfo {author} {\bibfnamefont {H.}~\bibnamefont {Rauf}}, \bibinfo {author}
  {\bibfnamefont {T.}~\bibnamefont {Pichler}}, \bibinfo {author} {\bibfnamefont
  {J.}~\bibnamefont {Bernardi}}, \bibinfo {author} {\bibfnamefont
  {H.}~\bibnamefont {Peterlik}}, \bibinfo {author} {\bibfnamefont
  {L.}~\bibnamefont {Korecz}}, \bibinfo {author} {\bibfnamefont
  {F.}~\bibnamefont {F\"ul\"op}}, \ and\ \bibinfo {author} {\bibfnamefont
  {A.}~\bibnamefont {J\'anossy}},\ }\href@noop {} {\bibfield  {journal}
  {\bibinfo  {journal} {Chemical Physics Letters}\ }\textbf {\bibinfo {volume}
  {383}},\ \bibinfo {pages} {362 } (\bibinfo {year} {2004})}\BibitemShut
  {NoStop}%
\bibitem [{\citenamefont {Dresselhaus}\ and\ \citenamefont
  {Dresselhaus}(1981)}]{Dresselhaus1981}%
  \BibitemOpen
  \bibfield  {author} {\bibinfo {author} {\bibfnamefont {M.~S.}\ \bibnamefont
  {Dresselhaus}}\ and\ \bibinfo {author} {\bibfnamefont {G.}~\bibnamefont
  {Dresselhaus}},\ }\href@noop {} {\bibfield  {journal} {\bibinfo  {journal}
  {Advances in Physics}\ }\textbf {\bibinfo {volume} {30}},\ \bibinfo {pages}
  {1} (\bibinfo {year} {1981})}\BibitemShut {NoStop}%
\bibitem [{\citenamefont {F\'abi\'an}\ \emph {et~al.}(2012)\citenamefont
  {F\'abi\'an}, \citenamefont {D\'ora}, \citenamefont {Antal}, \citenamefont
  {Szolnoki}, \citenamefont {Korecz}, \citenamefont {Rockenbauer},
  \citenamefont {Nemes}, \citenamefont {Forr\'o},\ and\ \citenamefont
  {Simon}}]{FabianPRB2012}%
  \BibitemOpen
  \bibfield  {author} {\bibinfo {author} {\bibfnamefont {G.}~\bibnamefont
  {F\'abi\'an}}, \bibinfo {author} {\bibfnamefont {B.}~\bibnamefont {D\'ora}},
  \bibinfo {author} {\bibfnamefont {A.}~\bibnamefont {Antal}}, \bibinfo
  {author} {\bibfnamefont {L.}~\bibnamefont {Szolnoki}}, \bibinfo {author}
  {\bibfnamefont {L.}~\bibnamefont {Korecz}}, \bibinfo {author} {\bibfnamefont
  {A.}~\bibnamefont {Rockenbauer}}, \bibinfo {author} {\bibfnamefont {N.~M.}\
  \bibnamefont {Nemes}}, \bibinfo {author} {\bibfnamefont {L.}~\bibnamefont
  {Forr\'o}}, \ and\ \bibinfo {author} {\bibfnamefont {F.}~\bibnamefont
  {Simon}},\ }\href@noop {} {\bibfield  {journal} {\bibinfo  {journal}
  {Physical Review B}\ }\textbf {\bibinfo {volume} {85}},\ \bibinfo {pages}
  {235405} (\bibinfo {year} {2012})}\BibitemShut {NoStop}%
\bibitem [{\citenamefont {Donovan}\ \emph {et~al.}(1993)\citenamefont
  {Donovan}, \citenamefont {Klein}, \citenamefont {Dressel}, \citenamefont
  {Holczer},\ and\ \citenamefont {Gr\"uner}}]{Donovan1993}%
  \BibitemOpen
  \bibfield  {author} {\bibinfo {author} {\bibfnamefont {S.}~\bibnamefont
  {Donovan}}, \bibinfo {author} {\bibfnamefont {O.}~\bibnamefont {Klein}},
  \bibinfo {author} {\bibfnamefont {M.}~\bibnamefont {Dressel}}, \bibinfo
  {author} {\bibfnamefont {K.}~\bibnamefont {Holczer}}, \ and\ \bibinfo
  {author} {\bibfnamefont {G.}~\bibnamefont {Gr\"uner}},\ }\href@noop {}
  {\bibfield  {journal} {\bibinfo  {journal} {International Journal of Infrared
  and Millimeter Waves}\ }\textbf {\bibinfo {volume} {14}},\ \bibinfo {pages}
  {2459} (\bibinfo {year} {1993})}\BibitemShut {NoStop}%
\bibitem [{\citenamefont {Kitano}\ \emph
  {et~al.}(2002{\natexlab{c}})\citenamefont {Kitano}, \citenamefont {Matsuo},
  \citenamefont {Miwa}, \citenamefont {Maeda}, \citenamefont {Takenobu},
  \citenamefont {Iwasa},\ and\ \citenamefont {Mitani}}]{KitanoPRL2002}%
  \BibitemOpen
  \bibfield  {author} {\bibinfo {author} {\bibfnamefont {H.}~\bibnamefont
  {Kitano}}, \bibinfo {author} {\bibfnamefont {R.}~\bibnamefont {Matsuo}},
  \bibinfo {author} {\bibfnamefont {K.}~\bibnamefont {Miwa}}, \bibinfo {author}
  {\bibfnamefont {A.}~\bibnamefont {Maeda}}, \bibinfo {author} {\bibfnamefont
  {T.}~\bibnamefont {Takenobu}}, \bibinfo {author} {\bibfnamefont
  {Y.}~\bibnamefont {Iwasa}}, \ and\ \bibinfo {author} {\bibfnamefont
  {T.}~\bibnamefont {Mitani}},\ }\href@noop {} {\bibfield  {journal} {\bibinfo
  {journal} {Physical Review Letters}\ }\textbf {\bibinfo {volume} {88}},\
  \bibinfo {pages} {096401} (\bibinfo {year} {2002}{\natexlab{c}})}\BibitemShut
  {NoStop}%
\bibitem [{\citenamefont {Cs\H{o}sz}\ \emph {et~al.}(2018)\citenamefont
  {Cs\H{o}sz}, \citenamefont {M\'arkus}, \citenamefont {J\'anossy},
  \citenamefont {Nemes}, \citenamefont {Mur\'anyi}, \citenamefont {Klupp},
  \citenamefont {Kamar\'as}, \citenamefont {Kogan}, \citenamefont {Bud'ko},
  \citenamefont {Canfield},\ and\ \citenamefont {Simon}}]{Csosz2018}%
  \BibitemOpen
  \bibfield  {author} {\bibinfo {author} {\bibfnamefont {G.}~\bibnamefont
  {Cs\H{o}sz}}, \bibinfo {author} {\bibfnamefont {B.~G.}\ \bibnamefont
  {M\'arkus}}, \bibinfo {author} {\bibfnamefont {A.}~\bibnamefont {J\'anossy}},
  \bibinfo {author} {\bibfnamefont {N.~M.}\ \bibnamefont {Nemes}}, \bibinfo
  {author} {\bibfnamefont {F.}~\bibnamefont {Mur\'anyi}}, \bibinfo {author}
  {\bibfnamefont {G.}~\bibnamefont {Klupp}}, \bibinfo {author} {\bibfnamefont
  {K.}~\bibnamefont {Kamar\'as}}, \bibinfo {author} {\bibfnamefont {V.~G.}\
  \bibnamefont {Kogan}}, \bibinfo {author} {\bibfnamefont {S.~L.}\ \bibnamefont
  {Bud'ko}}, \bibinfo {author} {\bibfnamefont {P.~C.}\ \bibnamefont
  {Canfield}}, \ and\ \bibinfo {author} {\bibfnamefont {F.}~\bibnamefont
  {Simon}},\ }\href@noop {} {\bibfield  {journal} {\bibinfo  {journal}
  {Scientific Reports}\ }\textbf {\bibinfo {volume} {8}},\ \bibinfo {pages}
  {11480} (\bibinfo {year} {2018})}\BibitemShut {NoStop}%
\bibitem [{\citenamefont {Gy\"ure-Garami}\ \emph {et~al.}(2018)\citenamefont
  {Gy\"ure-Garami}, \citenamefont {S\'agi}, \citenamefont {M\'arkus},\ and\
  \citenamefont {Simon}}]{Gyure2018}%
  \BibitemOpen
  \bibfield  {author} {\bibinfo {author} {\bibfnamefont {B.}~\bibnamefont
  {Gy\"ure-Garami}}, \bibinfo {author} {\bibfnamefont {O.}~\bibnamefont
  {S\'agi}}, \bibinfo {author} {\bibfnamefont {B.~G.}\ \bibnamefont
  {M\'arkus}}, \ and\ \bibinfo {author} {\bibfnamefont {F.}~\bibnamefont
  {Simon}},\ }\href@noop {} {\bibfield  {journal} {\bibinfo  {journal}
  {Available on arXiv:1805.11347}\ } (\bibinfo {year} {2018})}\BibitemShut
  {NoStop}%
\bibitem [{\citenamefont {Karsa}\ \emph {et~al.}(2012)\citenamefont {Karsa},
  \citenamefont {Quintavalle}, \citenamefont {Forr\'o},\ and\ \citenamefont
  {Simon}}]{KarsaPssb2012}%
  \BibitemOpen
  \bibfield  {author} {\bibinfo {author} {\bibfnamefont {A.}~\bibnamefont
  {Karsa}}, \bibinfo {author} {\bibfnamefont {D.}~\bibnamefont {Quintavalle}},
  \bibinfo {author} {\bibfnamefont {L.}~\bibnamefont {Forr\'o}}, \ and\
  \bibinfo {author} {\bibfnamefont {F.}~\bibnamefont {Simon}},\ }\href@noop {}
  {\bibfield  {journal} {\bibinfo  {journal} {physica status solidi (b)}\
  }\textbf {\bibinfo {volume} {249}},\ \bibinfo {pages} {2487} (\bibinfo {year}
  {2012})}\BibitemShut {NoStop}%
\bibitem [{\citenamefont {Klein}\ \emph
  {et~al.}(1994{\natexlab{b}})\citenamefont {Klein}, \citenamefont {Nicol},
  \citenamefont {Holczer},\ and\ \citenamefont {Gr\"uner}}]{HolczerPRB1994}%
  \BibitemOpen
  \bibfield  {author} {\bibinfo {author} {\bibfnamefont {O.}~\bibnamefont
  {Klein}}, \bibinfo {author} {\bibfnamefont {E.~J.}\ \bibnamefont {Nicol}},
  \bibinfo {author} {\bibfnamefont {K.}~\bibnamefont {Holczer}}, \ and\
  \bibinfo {author} {\bibfnamefont {G.}~\bibnamefont {Gr\"uner}},\ }\href@noop
  {} {\bibfield  {journal} {\bibinfo  {journal} {Physical Review B}\ }\textbf
  {\bibinfo {volume} {50}},\ \bibinfo {pages} {6307} (\bibinfo {year}
  {1994}{\natexlab{b}})}\BibitemShut {NoStop}%
\bibitem [{\citenamefont {M\'arkus}\ \emph {et~al.}(2018)\citenamefont
  {M\'arkus}, \citenamefont {Lloret}, \citenamefont {Wild}, \citenamefont
  {Domanov}, \citenamefont {Eisterer}, \citenamefont {Hauke}, \citenamefont
  {Pichler}, \citenamefont {Simon}, \citenamefont {Abell\'an}, \citenamefont
  {Shiozawa},\ and\ \citenamefont {Hirsch}}]{Shiozawa2018}%
  \BibitemOpen
  \bibfield  {author} {\bibinfo {author} {\bibfnamefont {B.~G.}\ \bibnamefont
  {M\'arkus}}, \bibinfo {author} {\bibfnamefont {V.}~\bibnamefont {Lloret}},
  \bibinfo {author} {\bibfnamefont {S.}~\bibnamefont {Wild}}, \bibinfo {author}
  {\bibfnamefont {O.}~\bibnamefont {Domanov}}, \bibinfo {author} {\bibfnamefont
  {M.}~\bibnamefont {Eisterer}}, \bibinfo {author} {\bibfnamefont
  {F.}~\bibnamefont {Hauke}}, \bibinfo {author} {\bibfnamefont
  {T.}~\bibnamefont {Pichler}}, \bibinfo {author} {\bibfnamefont
  {F.}~\bibnamefont {Simon}}, \bibinfo {author} {\bibfnamefont
  {G.}~\bibnamefont {Abell\'an}}, \bibinfo {author} {\bibfnamefont
  {H.}~\bibnamefont {Shiozawa}}, \ and\ \bibinfo {author} {\bibfnamefont
  {A.}~\bibnamefont {Hirsch}},\ }\href@noop {} {\bibfield  {journal} {\bibinfo
  {journal} {submitted, arXiv link will be available soon}\ } (\bibinfo {year}
  {2018})}\BibitemShut {NoStop}%
\bibitem [{\citenamefont {Zhang}\ \emph {et~al.}(2017)\citenamefont {Zhang},
  \citenamefont {Waters}, \citenamefont {Geim},\ and\ \citenamefont
  {Grigorieva}}]{ZhangNatComm2017}%
  \BibitemOpen
  \bibfield  {author} {\bibinfo {author} {\bibfnamefont {R.}~\bibnamefont
  {Zhang}}, \bibinfo {author} {\bibfnamefont {J.}~\bibnamefont {Waters}},
  \bibinfo {author} {\bibfnamefont {A.~K.}\ \bibnamefont {Geim}}, \ and\
  \bibinfo {author} {\bibfnamefont {I.~V.}\ \bibnamefont {Grigorieva}},\
  }\href@noop {} {\bibfield  {journal} {\bibinfo  {journal} {Nature
  Communications}\ }\textbf {\bibinfo {volume} {8}},\ \bibinfo {pages} {15036}
  (\bibinfo {year} {2017})}\BibitemShut {NoStop}%
\bibitem [{\citenamefont {M\'arkus}\ \emph {et~al.}(2017)\citenamefont
  {M\'arkus}, \citenamefont {Simon}, \citenamefont {Nagy}, \citenamefont
  {Feh\'er}, \citenamefont {Wild}, \citenamefont {Abell\'an}, \citenamefont
  {Chac\'on‐Torres}, \citenamefont {Hirsch},\ and\ \citenamefont
  {Hauke}}]{MarkusPssb2017}%
  \BibitemOpen
  \bibfield  {author} {\bibinfo {author} {\bibfnamefont {B.~G.}\ \bibnamefont
  {M\'arkus}}, \bibinfo {author} {\bibfnamefont {F.}~\bibnamefont {Simon}},
  \bibinfo {author} {\bibfnamefont {K.}~\bibnamefont {Nagy}}, \bibinfo {author}
  {\bibfnamefont {T.}~\bibnamefont {Feh\'er}}, \bibinfo {author} {\bibfnamefont
  {S.}~\bibnamefont {Wild}}, \bibinfo {author} {\bibfnamefont {G.}~\bibnamefont
  {Abell\'an}}, \bibinfo {author} {\bibfnamefont {J.~C.}\ \bibnamefont
  {Chac\'on‐Torres}}, \bibinfo {author} {\bibfnamefont {A.}~\bibnamefont
  {Hirsch}}, \ and\ \bibinfo {author} {\bibfnamefont {F.}~\bibnamefont
  {Hauke}},\ }\href@noop {} {\bibfield  {journal} {\bibinfo  {journal} {physica
  status solidi (b)}\ }\textbf {\bibinfo {volume} {254}},\ \bibinfo {pages}
  {1700232} (\bibinfo {year} {2017})}\BibitemShut {NoStop}%
\bibitem [{\citenamefont {Narita}\ \emph {et~al.}(1983)\citenamefont {Narita},
  \citenamefont {Akahama}, \citenamefont {Tsukiyama}, \citenamefont {Muro},
  \citenamefont {Mori}, \citenamefont {Endo}, \citenamefont {Taniguchi},
  \citenamefont {Seki}, \citenamefont {Suga}, \citenamefont {Mikuni},\ and\
  \citenamefont {Kanzaki}}]{Narita1983}%
  \BibitemOpen
  \bibfield  {author} {\bibinfo {author} {\bibfnamefont {S.}~\bibnamefont
  {Narita}}, \bibinfo {author} {\bibfnamefont {Y.}~\bibnamefont {Akahama}},
  \bibinfo {author} {\bibfnamefont {Y.}~\bibnamefont {Tsukiyama}}, \bibinfo
  {author} {\bibfnamefont {K.}~\bibnamefont {Muro}}, \bibinfo {author}
  {\bibfnamefont {S.}~\bibnamefont {Mori}}, \bibinfo {author} {\bibfnamefont
  {S.}~\bibnamefont {Endo}}, \bibinfo {author} {\bibfnamefont {M.}~\bibnamefont
  {Taniguchi}}, \bibinfo {author} {\bibfnamefont {M.}~\bibnamefont {Seki}},
  \bibinfo {author} {\bibfnamefont {S.}~\bibnamefont {Suga}}, \bibinfo {author}
  {\bibfnamefont {A.}~\bibnamefont {Mikuni}}, \ and\ \bibinfo {author}
  {\bibfnamefont {H.}~\bibnamefont {Kanzaki}},\ }\href@noop {} {\bibfield
  {journal} {\bibinfo  {journal} {Physica B+C}\ }\textbf {\bibinfo {volume}
  {117}},\ \bibinfo {pages} {422} (\bibinfo {year} {1983})}\BibitemShut
  {NoStop}%
\bibitem [{\citenamefont {Baba}\ \emph {et~al.}(1991)\citenamefont {Baba},
  \citenamefont {Izumida}, \citenamefont {Morita}, \citenamefont {Korke},\ and\
  \citenamefont {Fukase}}]{Baba1991_1}%
  \BibitemOpen
  \bibfield  {author} {\bibinfo {author} {\bibfnamefont {M.}~\bibnamefont
  {Baba}}, \bibinfo {author} {\bibfnamefont {F.}~\bibnamefont {Izumida}},
  \bibinfo {author} {\bibfnamefont {A.}~\bibnamefont {Morita}}, \bibinfo
  {author} {\bibfnamefont {Y.}~\bibnamefont {Korke}}, \ and\ \bibinfo {author}
  {\bibfnamefont {T.}~\bibnamefont {Fukase}},\ }\href@noop {} {\bibfield
  {journal} {\bibinfo  {journal} {Japanese Journal of Applied Physics}\
  }\textbf {\bibinfo {volume} {30}},\ \bibinfo {pages} {1753} (\bibinfo {year}
  {1991})}\BibitemShut {NoStop}%
\bibitem [{\citenamefont {Coffey}\ and\ \citenamefont
  {Clem}(1991)}]{CoffeyClemPRL1991}%
  \BibitemOpen
  \bibfield  {author} {\bibinfo {author} {\bibfnamefont {M.~W.}\ \bibnamefont
  {Coffey}}\ and\ \bibinfo {author} {\bibfnamefont {J.~R.}\ \bibnamefont
  {Clem}},\ }\href@noop {} {\bibfield  {journal} {\bibinfo  {journal} {Physical
  Review Letters}\ }\textbf {\bibinfo {volume} {67}},\ \bibinfo {pages} {386}
  (\bibinfo {year} {1991})}\BibitemShut {NoStop}%
\bibitem [{\citenamefont {Coffey}\ and\ \citenamefont
  {Clem}(1992{\natexlab{a}})}]{CoffeyClemPRB1992}%
  \BibitemOpen
  \bibfield  {author} {\bibinfo {author} {\bibfnamefont {M.~W.}\ \bibnamefont
  {Coffey}}\ and\ \bibinfo {author} {\bibfnamefont {J.~R.}\ \bibnamefont
  {Clem}},\ }\href@noop {} {\bibfield  {journal} {\bibinfo  {journal} {Physical
  Review B}\ }\textbf {\bibinfo {volume} {45}},\ \bibinfo {pages} {9872}
  (\bibinfo {year} {1992}{\natexlab{a}})}\BibitemShut {NoStop}%
\bibitem [{\citenamefont {Coffey}\ and\ \citenamefont
  {Clem}(1992{\natexlab{b}})}]{CoffeyClemPRB19922}%
  \BibitemOpen
  \bibfield  {author} {\bibinfo {author} {\bibfnamefont {M.~W.}\ \bibnamefont
  {Coffey}}\ and\ \bibinfo {author} {\bibfnamefont {J.~R.}\ \bibnamefont
  {Clem}},\ }\href@noop {} {\bibfield  {journal} {\bibinfo  {journal} {Physical
  Review B}\ }\textbf {\bibinfo {volume} {45}},\ \bibinfo {pages} {10527}
  (\bibinfo {year} {1992}{\natexlab{b}})}\BibitemShut {NoStop}%
\bibitem [{\citenamefont {Coffey}\ and\ \citenamefont
  {Clem}(1992{\natexlab{c}})}]{CoffeyClem19924}%
  \BibitemOpen
  \bibfield  {author} {\bibinfo {author} {\bibfnamefont {M.~W.}\ \bibnamefont
  {Coffey}}\ and\ \bibinfo {author} {\bibfnamefont {J.~R.}\ \bibnamefont
  {Clem}},\ }\href@noop {} {\bibfield  {journal} {\bibinfo  {journal} {Physical
  Review B}\ }\textbf {\bibinfo {volume} {46}},\ \bibinfo {pages} {11757}
  (\bibinfo {year} {1992}{\natexlab{c}})}\BibitemShut {NoStop}%
\bibitem [{\citenamefont {Coffey}\ and\ \citenamefont
  {Clem}(1993)}]{CoffeyClem1993}%
  \BibitemOpen
  \bibfield  {author} {\bibinfo {author} {\bibfnamefont {M.~W.}\ \bibnamefont
  {Coffey}}\ and\ \bibinfo {author} {\bibfnamefont {J.~R.}\ \bibnamefont
  {Clem}},\ }\href@noop {} {\bibfield  {journal} {\bibinfo  {journal} {Physical
  Review B}\ }\textbf {\bibinfo {volume} {48}},\ \bibinfo {pages} {342}
  (\bibinfo {year} {1993})}\BibitemShut {NoStop}%
\bibitem [{\citenamefont {Prassides}(2012)}]{PrassidesBook}%
  \BibitemOpen
  \bibfield  {author} {\bibinfo {author} {\bibfnamefont {K.}~\bibnamefont
  {Prassides}},\ }\href@noop {} {\emph {\bibinfo {title} {{Physics and
  Chemistry of the Fullerenes}}}}\ (\bibinfo  {publisher} {Springer-Verlag},\
  \bibinfo {address} {Berlin, Heidelberg},\ \bibinfo {year} {2012})\BibitemShut
  {NoStop}%
\bibitem [{\citenamefont {Gunnarsson}(1997)}]{GunnarsonRMP1997}%
  \BibitemOpen
  \bibfield  {author} {\bibinfo {author} {\bibfnamefont {O.}~\bibnamefont
  {Gunnarsson}},\ }\href@noop {} {\bibfield  {journal} {\bibinfo  {journal}
  {Reviews of Modern Physics}\ }\textbf {\bibinfo {volume} {69}},\ \bibinfo
  {pages} {575} (\bibinfo {year} {1997})}\BibitemShut {NoStop}%
\bibitem [{\citenamefont {Kaiser}\ \emph {et~al.}(1998)\citenamefont {Kaiser},
  \citenamefont {D\"usberg},\ and\ \citenamefont {Roth}}]{KaiserPRB1998}%
  \BibitemOpen
  \bibfield  {author} {\bibinfo {author} {\bibfnamefont {A.~B.}\ \bibnamefont
  {Kaiser}}, \bibinfo {author} {\bibfnamefont {G.}~\bibnamefont {D\"usberg}}, \
  and\ \bibinfo {author} {\bibfnamefont {S.}~\bibnamefont {Roth}},\ }\href@noop
  {} {\bibfield  {journal} {\bibinfo  {journal} {Physical Review B}\ }\textbf
  {\bibinfo {volume} {57}},\ \bibinfo {pages} {1418} (\bibinfo {year}
  {1998})}\BibitemShut {NoStop}%
\bibitem [{\citenamefont {Pichler}\ \emph {et~al.}(1999)\citenamefont
  {Pichler}, \citenamefont {Sing}, \citenamefont {Knupfer}, \citenamefont
  {Golden},\ and\ \citenamefont {Fink}}]{Pichler1999}%
  \BibitemOpen
  \bibfield  {author} {\bibinfo {author} {\bibfnamefont {T.}~\bibnamefont
  {Pichler}}, \bibinfo {author} {\bibfnamefont {M.}~\bibnamefont {Sing}},
  \bibinfo {author} {\bibfnamefont {M.}~\bibnamefont {Knupfer}}, \bibinfo
  {author} {\bibfnamefont {M.~S.}\ \bibnamefont {Golden}}, \ and\ \bibinfo
  {author} {\bibfnamefont {J.}~\bibnamefont {Fink}},\ }\href@noop {} {\bibfield
   {journal} {\bibinfo  {journal} {Solid State Communications}\ }\textbf
  {\bibinfo {volume} {109}},\ \bibinfo {pages} {721 } (\bibinfo {year}
  {1999})}\BibitemShut {NoStop}%
\end{thebibliography}%

\end{document}